\documentclass[12pt,preprint]{aastex}

\newcommand{\msunyr}{M$_{\sun}$~yr$^{-1}$~}
\newcommand{\numNBK}{139}
\newcommand{\numNBKnostars}{136}

\newcommand{\numNBJ}{122}
\newcommand{\numdual}{41}
\newcommand{\numNBKnoK}{13}
\slugcomment{Draft: \today}
\shorttitle{TBD}
\shortauthors{Lee et al.}

\begin{document}

\title{A Dual Narrowband Survey for H$\alpha$ Emitters at $z$=2.2:\\
Demonstration of the Technique and Constraints on the H$\alpha$ Luminosity Function\footnote{This paper is primarily based upon data gathered with the 6.5 meter Magellan Telescopes located at Las Campanas Observatory, Chile.}}

\author{
Janice C. Lee\altaffilmark{1,2,10,12},
Chun Ly\altaffilmark{2,11},
Lee Spitler\altaffilmark{4},
Ivo Labb\'{e}\altaffilmark{5,1},
Samir Salim\altaffilmark{6},\\
S. Eric Persson\altaffilmark{1},
Masami Ouchi\altaffilmark{7,8,1},
Daniel A. Dale\altaffilmark{9},
Andy Monson\altaffilmark{1},
David Murphy\altaffilmark{1}
}

\altaffiltext{1}{Carnegie Observatories, 813 Santa Barbara
Street, Pasadena, CA}
\altaffiltext{2}{Current address: Space Telescope Science Institute, Baltimore, MD}
\altaffiltext{3}{Visiting Astronomer, SSC/IPAC, Caltech, Pasadena, CA}
\altaffiltext{4}{Centre for Astrophysics and Supercomputing, Swinburne University of Technology, Australia}
\altaffiltext{5}{Leiden Observatory, Leiden University, Leiden, The Netherlands}
\altaffiltext{6}{Astronomy Department, Indiana University, Bloomington, IN}
\altaffiltext{7}{Institute for the Physics and Mathematics of the Universe (IPMU), TODIAS, The University of Tokyo, Japan}
\altaffiltext{8}{Institute for Cosmic Ray Research, The University of Tokyo, Japan}
\altaffiltext{9}{Department of Physics and Astronomy, University of Wyoming, Laramie, WY}
\altaffiltext{10}{Carnegie Fellow}
\altaffiltext{11}{Giacconi Fellow}
\altaffiltext{12}{jlee@stsci.edu}

\begin{abstract}
We present first results from a narrowband imaging program
for intermediate redshift emission-line galaxies
using the newly commissioned FourStar infrared camera 
at the 6.5m Magellan telescope.  To enable prompt
identification of H$\alpha$ emitters, 
a pair of custom 1\% filters,
which sample low-airglow atmospheric windows at 1.19 $\mu$m 
and 2.10 $\mu$m, is used to detect both H$\alpha$ 
and [OII]$\lambda$3727 emission 
from the same redshift volume at $z$=2.2.  
Initial observations are taken over a 130 arcmin$^2$ area in the 
CANDELS-COSMOS field.
The exquisite image quality resulting from the combination of the
instrument, telescope, and standard site conditions 
($\sim$0\farcs55 FWHM) 
allows the 1.19 $\mu$m and 2.10 $\mu$m data 
to probe 3$\sigma$ emission-line depths down to 
1.0$\times$10$^{-17}$ erg s$^{-1}$ cm$^{-2}$ 
and 1.2$\times$10$^{-17}$ erg s$^{-1}$ cm$^{-2}$
respectively, in less than 10 hours of integration time
in each narrowband.  For H$\alpha$ at $z=0.8$ and $z=2.2$,
these fluxes correspond to observed star formation rates of
$\sim$0.3 and $\sim$4  M$_{\odot}$ yr$^{-1}$, respectively.
We find \numNBJ~ sources with a 1.19 $\mu$m excess, and 
\numNBKnostars~ with a 2.10 $\mu$m excess,
\numdual~ of which show an excess in both bands.  
The dual narrowband technique, as implemented
here, is estimated to identify $\gtrsim$80\% 
of $z=2.2$ H$\alpha$ emitters in the narrowband excess population.
With the most secure such sample obtained to-date,
we compute constraints on the faint-end slope
of the $z=2.2$ H$\alpha$ luminosity function.  Fitting of a pure 
power-law gives $\alpha=-1.85\pm$0.31, which is steeper than other recent
estimates based on coarser selection techniques, but consistent
within the typically large uncertainties that currently characterize such
measurements.  Combining our LF points with those
at higher luminosities from other work, the slope decreases to
$\alpha=-1.58\pm0.40$.     
  These ``narrow-deep" FourStar
observations have been obtained as part
of the larger NewH$\alpha$ Survey, which will
combine the data with ``wide-shallow" imaging through a similar
narrowband filter pair with 
NEWFIRM at the KPNO/CTIO 4m telescopes, to enable
 study of both luminous (but rare) 
and faint emission-line
galaxies in the intermediate redshift universe. 

 
\end{abstract}





\section{Introduction}

The rest-frame optical emission lines are indispensable diagnostics of the activity in galaxies.  In the local Universe, the measurement of lines from [OII]$\lambda$3727 through [SII]$\lambda\lambda6716,31$ has traditionally provided essential information on properties such as the star formation rate (SFR); the physical state of the ISM (e.g., electron temperatures and densities, gas-phase metallicities); the dust reddening; and may also indicate whether an AGN is present.  Thus, it may seem like a conspiracy of nature that for $1<z<3$,  where 
star formation and AGN activity reach their global maxima (Shapley 2011 and references therein) and observing these lines is critical to the study of galaxy evolution, they are 
shifted into the near-infrared, where spectroscopic observations from the ground are difficult due to the bright sky background.

One way to partially circumvent the difficulties posed by the forest of near-infrared terrestrial lines is to use narrowband filters which are designed to 
transmit in dark regions of the sky spectrum.  Such filters can be employed to blindly detect and measure emission lines at select redshifts.  
In the past, this technique has been successfully used with red-sensitive optical imagers to search for Ly$\alpha$ emitters at the highest redshifts (e.g., Taniguchi et al. 2003 and references therein).  However, the near-IR instrumentation required to enable analogous surveys for galaxies via their rest-frame optical emission lines at ``cosmic high noon'' became readily available only around the mid-2000's.  Initial attempts with the first generation of IR imagers with small fields-of-view were heroic, but had limited success.  For example, Teplitz et al. (1998) collected a total of $\sim$48 hours of Keck observations with NIRC (38\arcsec\ FoV), and found a sample of just 13 candidates.  Even up to 2007, only a couple of hundred measurements of H$\alpha$ in field galaxies past $z$$\sim$ 0.4
had been published (e.g., Tresse et al. 2002; Erb et al. 2006c).
With the gradual maturation of wide-field near-infrared cameras however, efforts by a few independent groups (Villar et al. 2008; Geach et al. 2008; Sobral et al. 2009; Hayes et al. 2010), including our own (Ly et al. 2011; J.C. Lee et al. 2012, in preparation), have now succeeded in increasing 
the total number of candidate H$\alpha$ emitters at $z>0.8$ by an order of magnitude. 


Nevertheless, while using strategically designed filters allows for deep samples of emission-line galaxies at intermediate redshifts to be obtained, 
additional data are still needed
to identify the line responsible for the detected narrowband photometric excess.
Unambiguous determinations can be made with spectroscopy. However,
since spectra are not usually immediately available for
the selected sources, other strategies, typically based
on broadband photometry and colors, are first 
used to deduce the nature of the excess (e.g., Fujita et al. 2003;
Kodama et al. 2004; Ly et al. 2007; Shioya et al. 2008; Sobral et al. 2012).  
For surveys targeting intermediate redshift emission-line 
galaxies, such coarse filtering may be the only
method available to discriminate between different line emitters,
since a prohibitive investment of telescope time is still required to
obtain near-IR spectroscopy for  
large and deep samples (e.g., Geach et al. 2008).
The resulting contamination rate and incompleteness will vary
depending on the depth of the
photometry and number of colors available
(e.g., 15-30\% contamination if $Riz$ selection is used to select
H$\alpha$ emitters at $z=0.8$; Ly et al. 2011),
but will be most severe for young, blue galaxies, 
since the primary spectral feature providing information
on the approximate redshift, the Balmer/4000 \AA\ break, is
weak or non-existent in such populations. 

To address such issues, we have attempted to improve upon the basic narrowband technique.  Specifically, we wish to enable the prompt identification of H$\alpha$ emitting galaxies at $z\sim2$, and also provide an initial glimpse into the state of the ISM which is encoded into the rest-frame optical emission-lines.  Rather than use a single narrowband filter, we have designed a pair of filters, both of which sample dark regions of the near-IR sky spectrum, and are coupled to capture H$\alpha$ and [OII]$\lambda$3727 over the same redshift volume.  Emission-line galaxy candidates are selected
through detection of a photometric excess in a filter at 2.10 $\mu$m,
and H$\alpha$ emitters are then isolated through
the presence of a second narrowband excess at 1.19 $\mu$m.\footnote{This [OII]/H$\alpha$ dual narrowband selection technique has just recently been also used by Sobral et al. (2012) to obtain a galaxy sample at $z=1.47$.  There, [OII] and H$\alpha$ are detected using Subaru/Suprime-Cam imaging in NB921, and UKIRT/WFCAM imaging in ``$NB_H$" ($\lambda$=1.617 $\mu$m), respectively.}  
The H$\alpha$ line is of particular interest because 
its flux is a primary indicator of
the instantaneous star formation rate (SFR) -- the 
dust-corrected H$\alpha$ luminosity is directly proportional
to the ionizing flux. 
The [OII] flux has also been used as an SFR indicator, but its strength is more strongly dependent on the metallicity and the ionization state of the gas.\footnote{See Kennicutt (1998) and Kennicutt \& Evans (2012) for a review of star formation rate indicators.}  The observed [OII]/H$\alpha$ ratio thus can provide a coarse combined measure of the metallicity and reddening (e.g., Nakajima et al. 2012).  
Furthermore, this technique yields an excellent sample of emission-line galaxies at z=2.2 for follow-up NIR spectroscopy, as three key lines, [OII], and H$\alpha$ along with the flanking [NII] lines, are already pre-selected to fall in clean windows of the OH sky spectrum.  

Here, we describe the narrowband imaging in this filter pair
obtained by the NewH$\alpha$ Survey (J.C. Lee et al. 2012, in preparation) 
using the newly commissioned
FourStar infrared camera (S.E. Persson et al. 2012, in preparation)
at the 6.5 Magellan-Baade telescope.
Initial observations are carried out over an 11\arcmin$\times$11\arcmin\ area in the
Cosmic Evolution Survey field
(COSMOS; Scoville et al. 2007) which overlaps
HST WFC3/IR and ACS coverage by the Cosmic Assembly
Near-IR Deep Extragalactic Legacy Survey
(CANDELS; Grogin et al. 2011; Koekemoer et al. 2011), and WFC3/IR grism observations by the 3D-HST Survey (van Dokkum et al. 2011).
We take advantage of the wealth of 
deep, multi-wavelength imaging observations
available in COSMOS to construct spectral energy 
distributions (SEDs), and compute
photometric redshifts for the narrowband excess sources.
A unique aspect of the ensemble of data used in
the photometric redshift analysis is that it includes near-IR
medium-band imaging through five bands from the
FourStar Galaxy Evolution Survey (Z-FOURGE; I. Labb\'{e} et al. 2012, in preparation).
The Z-FOURGE filter set
provides a powerful complement to the NewH$\alpha$
filter pair, as it was designed to locate the Balmer- or
4000 \AA- breaks in galaxies at $1.5 \lesssim z \lesssim 3.5$, a
range in which a significant fraction of 2.10 $\mu$m narrowband selected
emission-line galaxies are expected to lie.  Together with
other public datasets, photometry in 
28 filters is used, spanning from the 
$U$ to $Ks$ bands ($\sim$3800 \AA\ to $\sim$2.2 $\mu$m). 
With the SEDs and photometric redshifts, we examine the
composition of the narrowband excess samples, and
test the effectiveness of the dual narrowband 
technique. 
 
The remainder of the paper is organized as follows.  In Section 2,
an overview of the NewH$\alpha$ and Z-FOURGE parent surveys is given,
and the FourStar observations and data reduction is summarized. 
The method for selecting narrowband excess samples and identifying dual narrowband
emitters is described in Section 3.  Section 4 discusses the construction 
of SEDs, 
and the computation of photometric redshifts.
An inventory is taken of the various detected line-emitters 
using the photometric redshifts and $BzK$ colors in Section 5,
and the effectiveness of the dual narrowband excess method for identifying 
a clean sample of $z=2.2$ H$\alpha$ emitters is evaluated.
In Section 6, this sample is 
compared with the rest-frame UV-selected $z\sim$2 galaxies of Erb et al. (2006a,b,c)
to ({\it i}) illustrate the parameter space probed by our observations relative to a 
well-studied sample of intermediate redshift star-forming galaxies; and ({\it ii})
test the applicability of local scaling relations used to remove the contribution of 
[NII]$\lambda\lambda$6548,83 to the narrowband excess flux, and correct for
dust attenuation.
Initial constraints on the $z=2.2$ H$\alpha$ luminosity function based
on our new observations are also presented.   
Finally, we summarize our findings, and describe future work with
the dataset in Section 7.

Throughout the paper, the standard $\Lambda$CDM cosmology 
with [$\Omega_{\Lambda}$, $\Omega_M$, $h_{70}$] = [0.7, 0.3, 1.0] is assumed.
Magnitudes are reported on the AB system.
Star formation rates are computed using the Kennicutt (1998)
prescription and hence assume a Salpeter IMF with mass limits of
0.1 and 100 M$_{\odot}$.

\section{Observations and Data Reduction}

\subsection{The New H$\alpha$ Survey}\label{sec:survey}

The FourStar narrowband imaging used in this analysis has been obtained
as part of the overall NewH$\alpha$ Survey, a program which
extends searches for emission-line galaxies into the intermediate
redshift universe (J.C. Lee et al. 2012, in preparation).  
The NewH$\alpha$ Survey has been designed to capture
statistical samples of both luminous (but rare), and
faint emission-line galaxies, by combining the near-infrared
imaging capabilities of NEWFIRM (Probst et al. 2008) 
at the KPNO/CTIO 4-m telescopes
(FoV 27\farcm6 x 27\farcm6) to cover large areas, and 
the exquisite image quality ($\sim$0\farcs5-0\farcs6 FWHM)
of FourStar at
the Las Campanas Magellan 6.5-m (FoV 10\farcm8 x 10\farcm8) to probe
luminosities that are a factor of three deeper over smaller fields.
Follow-up programs with GALEX, Spitzer and the Magellan-IMACS 
spectrograph are in various stages of progress as described in (J.C. Lee et al. 2012, in preparation) 
Narrowband observations at 1.19 $\mu$m with NEWFIRM are complete, and
results on (i) the H$\alpha$ luminosity function and SFR volume
density at $z=0.8$, and (ii) the dust attenuation of $z=0.8$ H$\alpha$ emitters
based on Balmer decrements measured from IMACS optical spectroscopy
are reported in Ly et al. (2011) and Momcheva et al. (2012), respectively. 

\begin{figure}
\epsscale{1.3}
\includegraphics[clip=true, width=2.4in, trim=0 120 170 70]{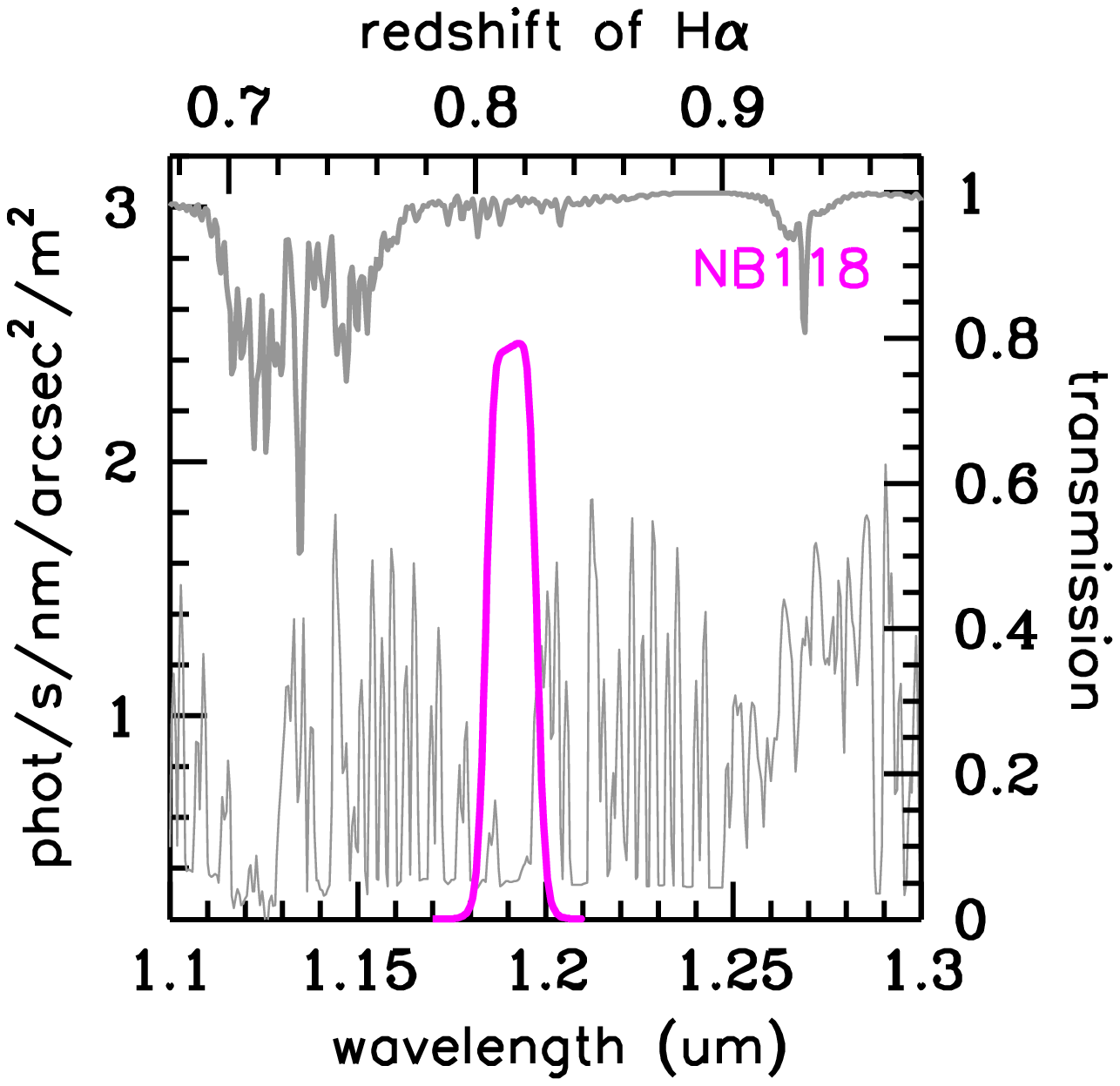}
\includegraphics[clip=true, width=2.4in, trim=0 120 170 70]{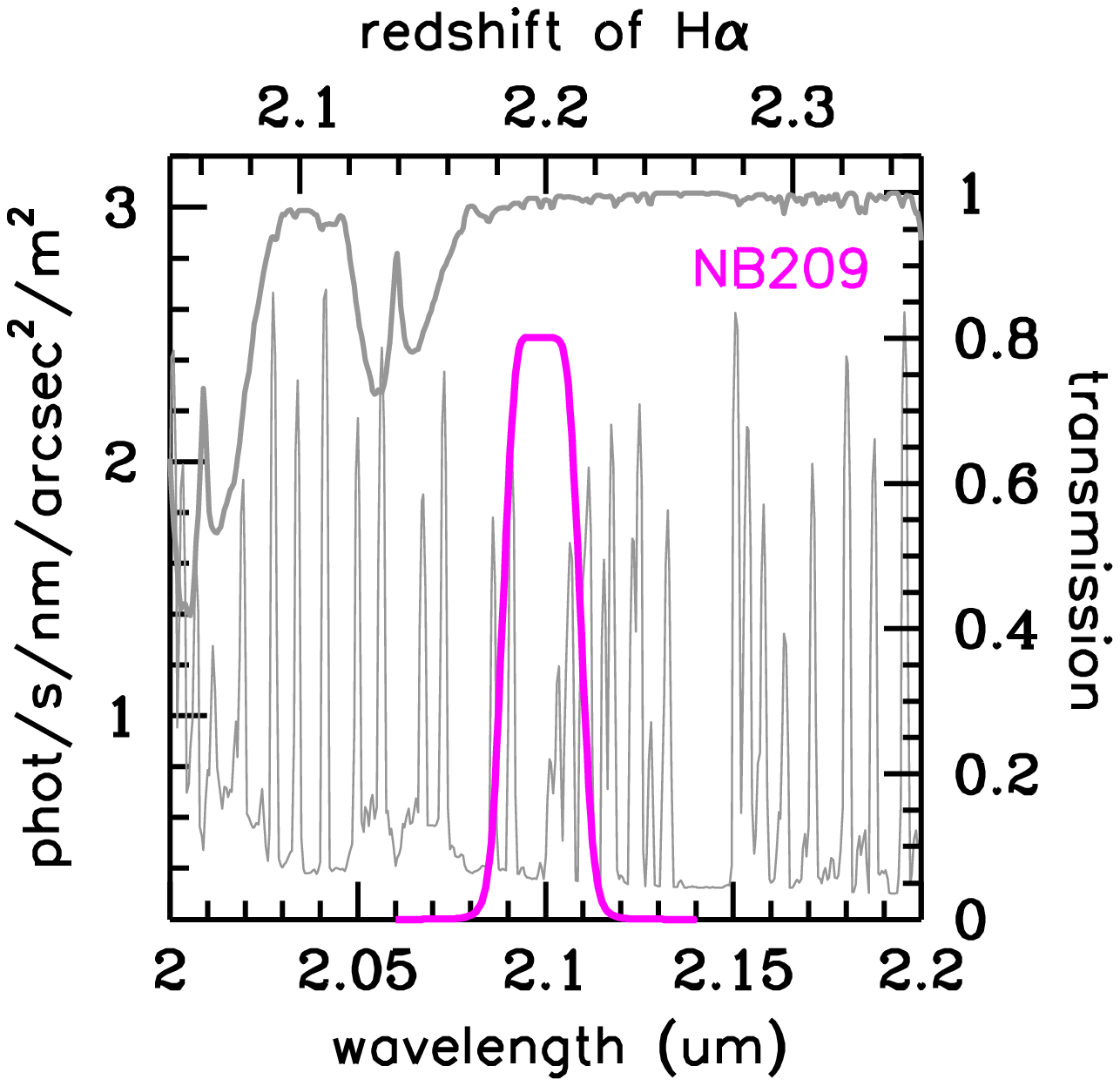}

\includegraphics[clip=true, width=5.2in, trim=0 100 0 0]{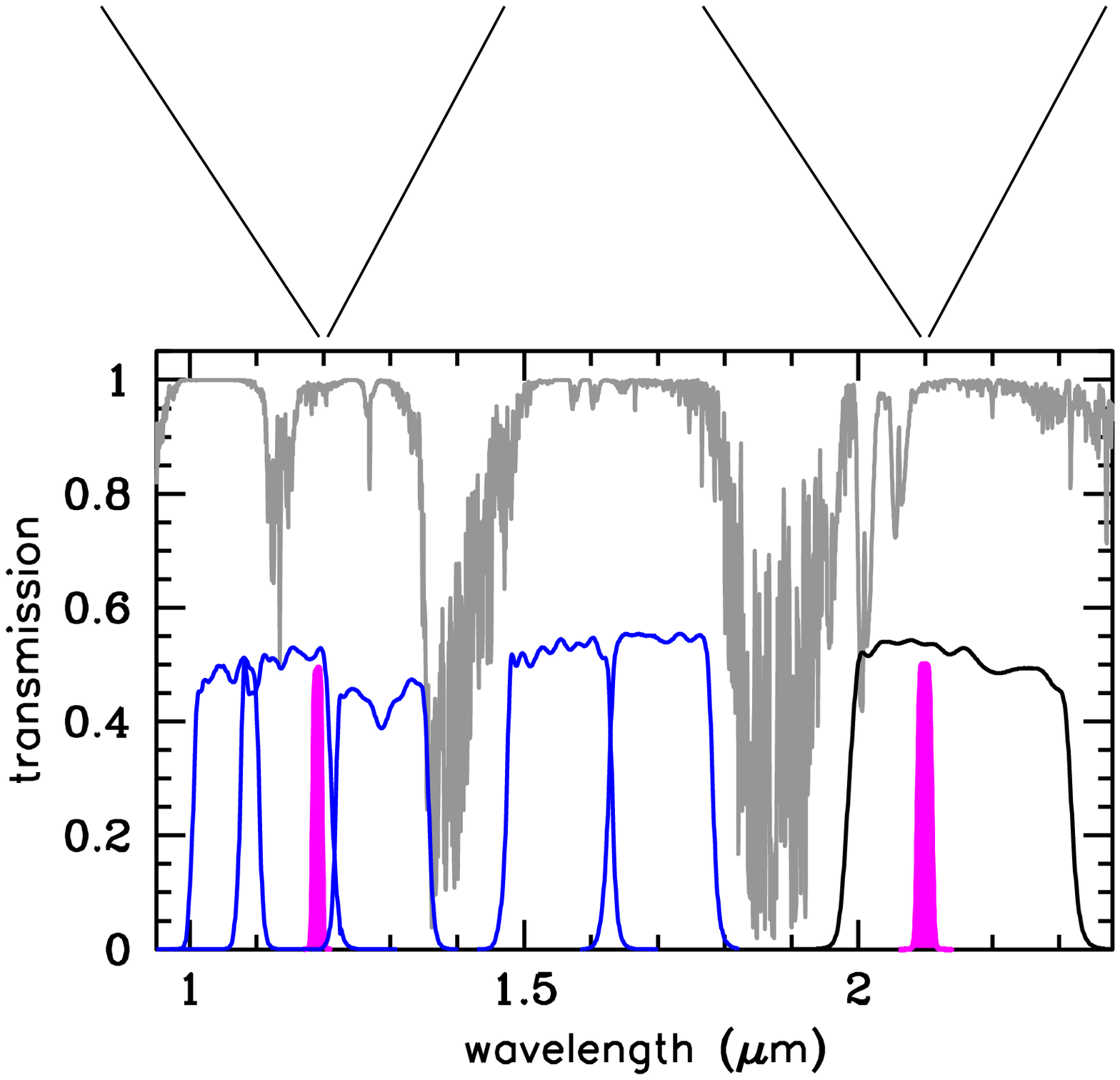}

\caption{Bandpasses of the NewH$\alpha$ narrowband filter pair (magenta) together with those for the Z-FOURGE medium-band filter set (blue), and the standard $Ks$ broadband (black), which are used in FourStar.  The atmospheric transmission is also plotted (gray).  The expanded view shows the profiles for the narrowbands overplotted on the OH sky spectrum (gray).  Similar sets of filters are also used in the NEWFIRM camera.  The narrowband filters probe low OH airglow regions in the sky spectrum, and are sensitive to H$\alpha$ at $z=0.8$ (near the beginning of the ten-fold decline in the cosmic SFR density) and at $z=2.2$ (within the peak of the cosmic star formation history). Sky spectrum and atmospheric transmission data from Gemini Observatories.  Filter transmission data (shown in the upper plots) from Barr Associates.  FourStar/Magellan system throughputs (shown in the lower plot) from S.E. Persson et al. 2012, in preparation.}
\label{fig:filters}
\end{figure}

\begin{figure}
\epsscale{0.5}
\plotone{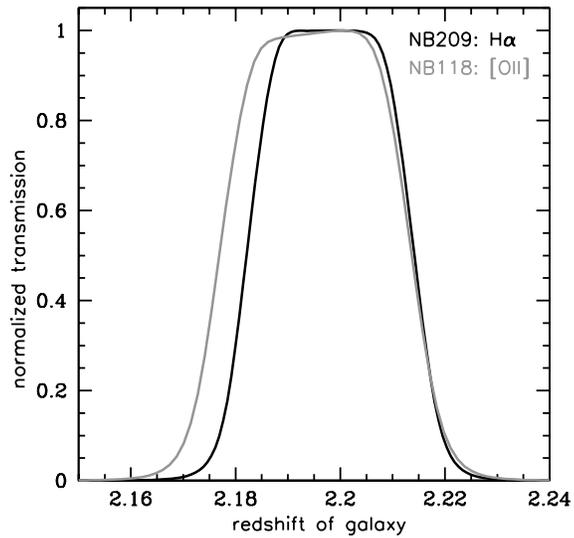}
\caption{The NewH$\alpha$ narrowband filters are coupled such that the [OII] emission 
of H$\alpha$ emitters selected with NB210 observations can be captured by the NB119 filter. 
 The redshift ranges over which the filters are sensitive to [OII] and H$\alpha$ in the same volume are illustrated in this Figure.  Bandpass shifting over the field-of-view of the detector is not significant because of the positioning of the filters near the focal plane in the camera.  The relative shift between the center and the corner of the field has been computed to be -6.10\AA\ for NB119, and -5.49\AA\ for NB210, which are 6\% and 2.6\% of the bandpass 
widths of the two filters.   
The redshift range over which NB119 is sensitive to [OII] 
completely encloses the range where NB210 will detect H$\alpha$.}
\label{fig:overlap}
\end{figure}

For both NEWFIRM and FourStar, a pair of custom narrowband filters, which
sample low OH-airglow windows in the near-IR sky spectrum, 
has been manufactured.  The bandpasses are made to fit within
high atmospheric transmission regions at 1.19 and 2.10 $\mu$m
which are relatively free of terrestrial sky lines over
1\% in $\lambda$ 
(Figure~\ref{fig:filters}, top).  
The filters are sensitive to H$\alpha$ at $z$=0.8
(near the beginning of the ten-fold decline in the
cosmic SFR density) 
and at $z$=2.2 (within the peak of the cosmic star
formation history). 
Fortuitously, filters centered
around these two wavelengths
can also be designed so that both [OII]$\lambda$3727
and H$\alpha$ emission can be detected from the same galaxies
at $z=2.2$ (Figure~\ref{fig:overlap}).  Again, 
such dual narrowband excess detections
help to differentiate H$\alpha$ emitters from other
emission-line sources prior to confirmation with
near-IR spectroscopy.  Another useful feature of
the particular redshift singled out
by this filter pair is that Ly$\alpha$ appears
at $\sim$3900 \AA, which is just above the
blue cut-off of most optical cameras.  Thus,
a third narrowband can be designed to also detect
Ly$\alpha$ in the same volumes, as
has been done by Nakajima et al. (2012) (see also Hayes et al. 2010).


The central wavelength
and FWHM of the two filters built for FourStar are 11908 \AA\ and 137 \AA\,
and 20987 \AA\ and 208 \AA.  Hereafter, we refer to these filters as NB119 and
NB210, respectively.

The observational goal of the NewH$\alpha$ survey's FourStar program is to probe 
line luminosities that are roughly an order of magnitude
fainter than the ``knee" of the H$\alpha$ luminosity 
function at $z=2.2$, to
allow for example, robust constraints on the faint end slope to be derived.  
Based on measurements made by Geach et al. (2008) and Hayes et al. 
(2010), this depth is estimated to be around H$\alpha$ luminosities
which translate into 
to observed SFRs of 5 \msunyr at $z\sim2.2$,
(fluxes of $\sim1.7\times10^{-17}$ ergs s$^{-1}$ cm$^{-2}$).  This limit can 
be reached in NB210 imaging with a limiting AB
magnitude of $\sim$23.7.  The corresponding depth required in the NB119 imaging
to detect [OII] for the faintest H$\alpha$ $z=2.2$ emitters is roughly
$8.5\times10^{-18}$ ergs s$^{-1}$ cm$^{-2}$ (25.2 AB), if it is assumed
that the observed H$\alpha$/[OII] ratio is 2.  The assumption is based
on spectroscopy of local star-forming galaxies.  Such studies show an average ratio
of $\sim$2, with values ranging between 1 and 4 for massive L$^*$ systems, and  
an average of $\sim$1.5 with values between 1 and 2 
for lower luminosity dwarf galaxies (Jansen et al. 2001; Moustakas et al. 2006; see also Figure~\ref{fig:oiiha}).
The effectiveness of these chosen relative depths for identifying
the $z=2.2$ H$\alpha$ emitters in the NB210 excess sample is examined in
Section~\ref{sec:composition}.

\subsection{Z-FOURGE}
Z-FOURGE, the FourStar Galaxy Evolution Survey (I. Labb\'{e} et al. 2012, in preparation),
is a program which aims to characterize 
the properties of galaxies at $z>1$ by performing
the first crucial (and often limiting) step of deriving reasonably
accurate redshifts (i.e., $\sim$2\% in $\frac{\Delta z}{1+z}$). 
Ultimately, Z-FOURGE seeks to measure such redshifts
for a sample of $\sim$30,000 $K_s$-selected galaxies at 
$1.5 \lesssim z \lesssim 3.5$.  The principal focus of Z-FOURGE 
is to investigate the red, early-type population using mass-selected
samples.  Since quiescent, non-active 
L$^*$ galaxies at
$z>1$ are too faint for spectroscopy, Z-FOURGE's strategy
is to use a custom set of five medium-band, near-infrared filters
to constrain the location of the Balmer/4000 \AA\ break.  
The filters sample wavelengths between 
1 to 1.8 $\mu$m, and splits the standard $J$ and $H$ bands into
three ($J1$, $J2$, $J3$) and two ($Hl$, $Hs$) bands, respectively.
A similar filter set has also been used with NEWFIRM by the
NEWFIRM Medium-Band Survey (van Dokkum et al 2009; Whitaker et al. 2011).

Bandpasses for the Z-FOURGE medium-band filter set are shown
together with the NewH$\alpha$ filter pair in Figure~\ref{fig:filters}.

\subsection{FourStar Observations}

Pilot study observations for the NewH$\alpha$ and Z-FOURGE
surveys were carried out during ``science demonstration time" in 
FourStar's first semester (2011A) on the Magellan-Baade 6.5m 
telescope.  Both programs targeted the same region 
centered on $\alpha=10^h00^m32^s.4$ $\delta=+02^{\circ}16\arcmin58\arcsec$,
within the CANDELS-COSMOS field.
Standard near-infrared observing procedures were followed.
The narrowband observations are summarized here.  A description of the 
medium-band observations can be found in the Z-FOURGE survey paper 
(I. Labb\'{e} et al. 2012, in preparation).

Narrowband observations were performed over a span of eight 
nights (24 January; 19, 21-22 February; 13-14, 19-20 March 2011),
which were shared with the Z-FOURGE survey.  Cumulative integration 
times of 9.65 and 8.39 hrs in the NB119 and NB210 filters
were obtained on a single FourStar
pointing.   The exposure times for individual NB119 and NB210
frames were 4 and 2 minutes, respectively.  The data were taken
with the instrumental gain in its low noise mode ($\sim$1.3 e-/ADU).
Semi-random dither patterns, where the spacing between
adjacent frames is maximized and the pointing centers are distributed over
a 27\arcsec\ box, are used.  The dithers serve to bridge the 18\arcsec\ 
gaps between FourStar's
four 2048 $\times$ 2048 HAWAII-2RG detectors; 
enable filtering and removal of bad pixels;
and perhaps most importantly,
provide a measurement of the sky in each pixel for accurate sky-subtraction.  
The median
seeing during the observations was 0\farcs6, and varied between
0\farcs45 and 0\farcs90.  Hence, the PSF was adequately
sampled by FourStar's 0\farcs159 pixels.  Sky conditions 
were mostly photometric,
although some data were taken through thin cirrus.
Continuum measurements are provided by imaging in the $K_s$ band
for the NB210 data, and by the $J2$ and $J3$ bands for NB119.

A summary of the observations in given in Table~\ref{tab:observations}.

\subsection{FourStar Data Reduction}

The FourStar data were reduced using a custom pipeline, adapted
from the IDL code used for the NEWFIRM Medium-Band Survey (see
Whitaker et al. 2011).
The reduction follows
standard, iterative (two-pass), near-infrared reduction 
techniques (e.g., Labbe et al. 2003)
to subtract the sky background; reject artifacts; optimally
weight and stack the individual frames; and is consistent 
with the procedures used to process the NewH$\alpha$ narrowband
data taken with NEWFIRM (Ly et al. 2011; J.C. Lee et al. 2012, in preparation). 
Corrections are applied for non-linearity and geometric distortion,
and flat-fielding is currently performed using twilight flats taken in each band.
Comparison with independent, publicly available, broadband near-IR 
imaging in COSMOS shows that large-scale systematic variations
in the sensitivity over the array have been reduced 
to $<$5\%.
Photometric calibration is performed using observations of 
the spectrophotometric standards GD71 and GD153, and the
resultant zeropoints are accurate to within $\sim$2\%.
Further optimization of the reduction and calibration is in progress.

The 3-$\sigma$ limiting depths of the stacked imaging
in a 1\farcs2 diameter aperture are given in Table~\ref{tab:observations}.
A 1\farcs2 diameter aperture is roughly twice the size of the PSF,
and will contain 99\% of light from a
point source, assuming a Gaussian profile.
It is the size of the aperture used to select 
narrowband excess sources, as described in the next section.
The narrowband observations successfully achieve the NB210 target depth described
in Section~\ref{sec:survey} (1.7$\times10^{-17}$ ergs s$^{-1}$ cm$^{-2}$) 
at the 5-$\sigma$ level, and surpass it
at the 3-$\sigma$ level.  The NB119 observations reach the
target depth at about the 3-$\sigma$ level.  Observations 
in the continuum bands reach flux limits
which are approximately twice as deep.
The limits are computed by measuring the background in a large number of random apertures.

\section{Sample Selection}

\subsection{Selection of NB210 Excess Objects}


Sources that show a significant $K_s-$NB210 color excess are selected
as emission-line galaxy candidates.  The selection procedure
follows general techniques commonly used in narrowband surveys
(e.g., Villar et al. 2008, Sobral. et al. 2009,
Ly et al. 2011, Nakajima et al. 2011).

First, we detect sources in the NB210 image with SExtractor,
using the default parameter values which require that 
detections have 5 adjacent pixels which are each individually above 1.2$\sigma$
of the background.  Exposure maps are used as weight images to
avoid spurious detections in low signal-to-noise regions.
A Gaussian convolution filter with a FWHM of 0\farcs6 
is applied to improve identification of faint, extended, objects.
A total of 4865 sources are found.
$K_s$ photometry is measured for each of
these NB210 detections in matched apertures using SExtractor
in ``dual-image" mode.  As discussed in the 
previous section, an aperture of  1\farcs2 diameter 
is chosen to compute the color.
 
\begin{figure}
\epsscale{1.1}
\plottwo{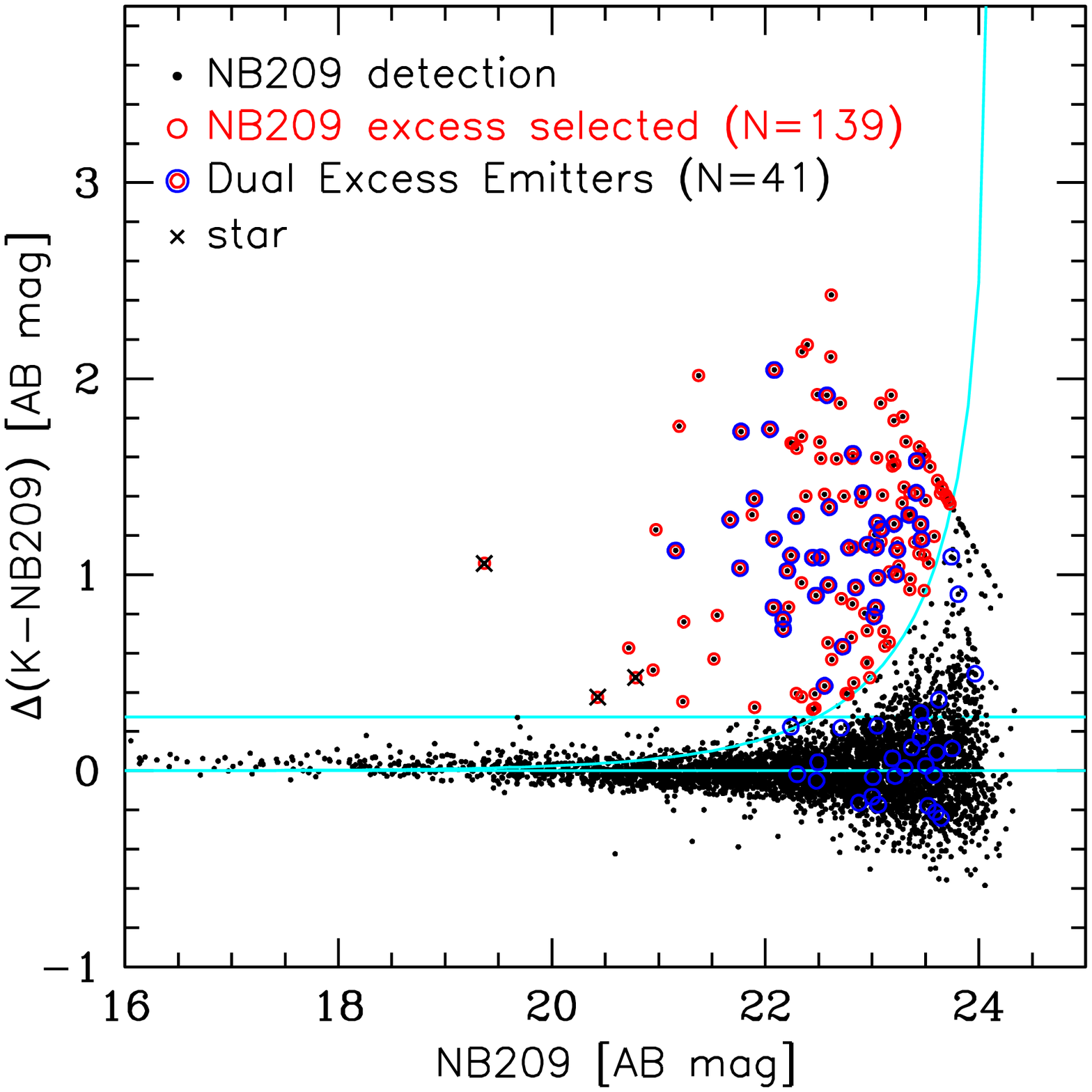}{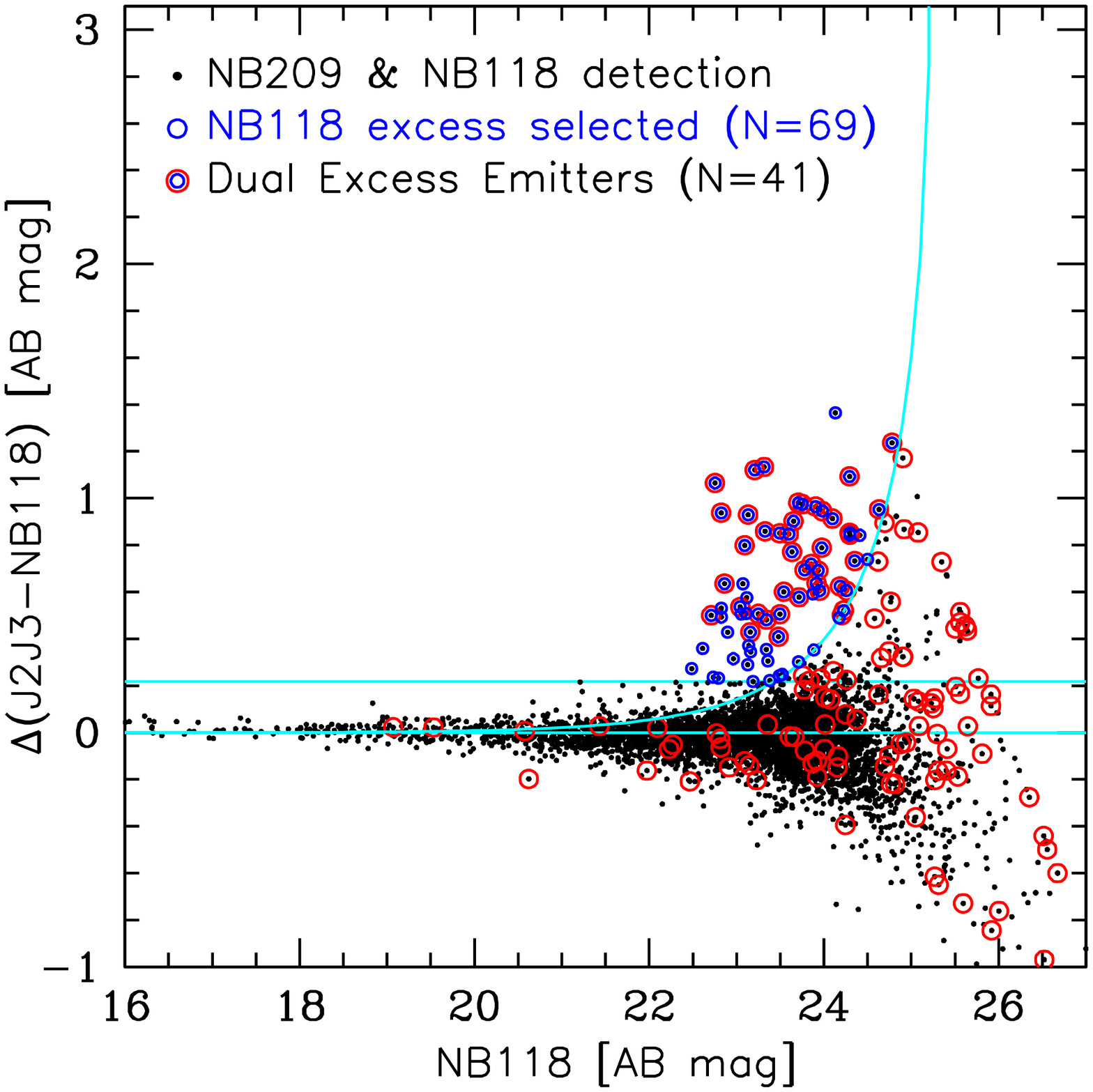}
\caption{Color-magnitude diagrams illustrating the selection of NB210 excess (left panel), 
and {\it NB210-detected}, NB119 excess (right panel) sources, 
where the photometry has been measured 
in 1\farcs2 diameter apertures.  The upper horizontal line indicates values which are five times 
the scatter in the color for continuum objects with magnitudes between 19.0 and 21.5. 
The curves indicate the values of the color excess that are significant at the
3-$\sigma$ level.  In both panels, the
NB119 excess sources are marked
with blue circles, and the NB210 excess sources with
red circles; hence the ``dual emitters" are marked in both
red and blue.  Three stars (black crosses) are identified via inspection of SEDs constructed 
from broadband photometry. 30\% of the NB210 excess sources are dual emitters.}
\label{fig:nb209cmd}
\end{figure}

Using this photometry, NB210 excess sources 
are then selected by requiring that: 

\noindent (i) $\Delta(K_s$-NB210), the $K_s$-NB210 color
corrected for the median color of bright continuum objects with $-0.4<K_s$-NB210$<0.4$
and $19.0<$NB210$<21.5$, has a positive value which is significant at
the 3$\sigma$ level or higher (curve in Figure~\ref{fig:nb209cmd}); and 

\noindent (ii)  $\Delta(K_s$-NB210) $>$ 0.275, which is five times the scatter in $\Delta(K_s$-NB210)
of the sample of continuum objects defined in (i) (upper horizontal line in Figure~\ref{fig:nb209cmd}).  The minimum color represents a threshold in the rest-frame equivalent width of H$\alpha$ at $z=2.2$ of 20 \AA\ (cf. Ly et al. 2011, eq. 11). 
 
The selection results in a sample of \numNBK~ objects, \numNBKnoK~ of which 
are undetected in the $K_s$-band. 
The objects are individually inspected on both the narrow and broad-band FourStar images, and none appear to be spurious detections.  SEDs constructed from broadband photometry (Section~\ref{sec:sedfitting}) are examined to identify stars.  Three stars are found and removed, reducing the NB210 excess sample to \numNBKnostars~ objects.

\subsection{Selection of Dual Narrowband Excess Emitters}

To determine which NB210 excess sources are ``dual emitters"
(i.e., they also show a photometric excess in the NB119 filter,
and are likely to be H$\alpha$ emitters at $z=2.2$),
we perform photometry on the NB119 image, and on a combined $J2$ and $J3$
(hereafter ``$J2J3$") image.  The measurements are again made with 
SExtractor in dual-image mode, with the detection performed on
the NB210 image.
This produces catalogs of NB119 and $J2J3$ photometry
at the locations of each of the 4865 NB210 detections.

Criteria analogous to those described in the previous section 
are used to select the NB119 excess sources from these catalogs.  
The corresponding color-magnitude diagram is given 
in the right panel of Figure~\ref{fig:nb209cmd}. 
The fraction of the NB210 excess sample which also 
show an excess in NB119 is 30\% (N=\numdual).  
In both panels of Figure~\ref{fig:nb209cmd}, the
NB119 excess sources are enclosed
with blue circles, and the NB210 excess sources with
red circles; hence the points marked in both red and
blue are dual emitters.  All of the dual 
emitters 
are detected in the $K_s$-band.

\subsection{Selection of NB119 Excess Objects}

\begin{figure}
\epsscale{0.5}
\plotone{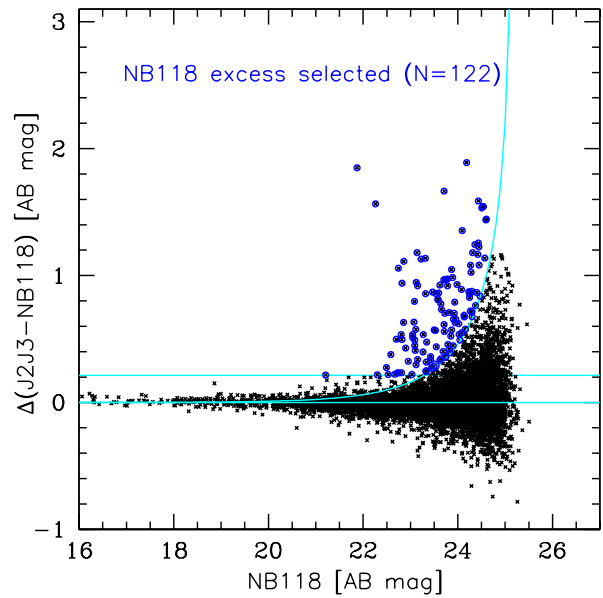}
\caption{Color-magnitude diagram illustrating the selection of a NB119 excess sample.  Whereas the NB119 color-magnitude diagram in Figure~\ref{fig:nb209cmd} shows photometry only at locations where there are NB210 detections, in this Figure, all sources detected on the NB119 image are plotted.  Again, the photometry has been measured in 1\farcs2 diameter apertures; the upper cyan horizontal line indicates values which are five times
the scatter in the color for continuum objects with magnitudes between 19.0 and 21.5; and the curve indicates values of the color excess that are significant at the
3-$\sigma$ level. }
\label{fig:nb118cmd}
\end{figure}

Following the above procedures 
source detection on the NB119 image and matched aperture photometry
on the $J2J3$ image can also be performed
to obtain an NB119 excess selected sample without reference
to the NB210 data. 
A total of 6894 sources are detected, with 
\numNBJ~ showing $\Delta(J2J3$-NB119)$>$0.21 
at at least 3$\sigma$ significance (Figure~\ref{fig:nb118cmd}).


\section{SEDs and Photometric Redshifts}\label{sec:sedfitting}

SEDs for the narrowband detected sources 
are constructed from the ensemble of publicly available 
optical broad-band and medium-band data
in the COSMOS field, in combination with the Magellan-Fourstar
NewH$\alpha$ and Z-FOURGE imaging described above.  
The SEDs are fit using the EAZY code (v1.1.10; Brammer et al. 2008)
to derive photometric redshifts.
We use the same optical dataset and a procedure similar to 
those in Whitaker et al. (2011), who performed a
near-IR medium-band study of the COSMOS field.
With EAZY, linear combination of seven template spectra,
which span a broad range of galaxy ages, along with a
Bayesian prior on the observed total $K_s$ magnitude, 
are used to to fit the observed SEDs.  
Based on these fits, a variety of different characteristic
photometric redshifts are calculated from the probability 
distribution of redshifts for a given
galaxy.  In the analysis that follows, ``$z_{peak}$," 
the redshift peak in the distribution
with the highest integrated probability,
is adopted as the photometric redshift.

The public datasets in the COSMOS field that are included in this analysis
consist of imaging in 22 filters from the $U$ to $Z$ bands ($\sim$3800 \AA\
to 9000 \AA).  We use data from: (1) the
Deep Canada-France-Hawaii Telescope
Legacy Survey  (CFHTLS $ugriz$, Erben et al. 2009; Hildebrandt et al. 2009), and
(2) Subaru Suprime-Cam ($BVr^{\prime}i^{\prime}z^{\prime}$, Taniguchi et al. 2007; and 12 medium-band
filters from 4200 \AA\ to 8200 \AA, Y. Taniguchi et al. 2012, in preparation).


Combining the COSMOS optical imaging in 22 filters with FourStar 
near-IR imaging in 5 Z-FOURGE medium-bands, 
and a standard $K_s$ filter, provides up to 
28 points for each individual SED.  
The imaging in the various bands
were all convolved to the same PSF (1\farcs24 FWHM),
and photometry was measured in 1\farcs5 diameter apertures. 
Details on the PSF matching of the imaging, and
the generation of a matched multi-wavelength flux catalog are
given in Spitler et al. (2012) and I. Labbe et al. (2012, in preparation). 


Our photometric redshifts are in reasonable agreement with available spectroscopic redshifts for the NB210 excess sources. 
We searched the zCOSMOS catalog (Lilly et al. 2007) for spectroscopic redshifts, 
and found 4 NB210 excess emitters with matches within 1\arcsec.  
All had $z_{spec}<1$, and none were dual excess emitters.  
Each of the 4 galaxies have individual
photometric and spectroscopic redshifts which are consistent to within 1\% in $\frac{\Delta z}{1+z_{spec}}$.
The NB210 excess for two galaxies can be identified as originating from
Pa$\alpha$ at $z=0.12$, and from HeI$\lambda$10830 at $z=0.94$ 
for the other two.  

An additional check of the accuracy is also performed by comparing the photo-z's of the larger parent sample of NB210 detections (as opposed to the NB210 {\it excess} detections) with available spectroscopic redshifts.
There are 458 galaxies with NB210 detections that have reliable spectroscopic redshifts (zCOSMOS ``confidence class" $>90\%$) within 1\arcsec.  
Comparison of the photometric and spectroscopic redshifts for $z_{spec}<1.4$ show a dispersion of $\sigma(\frac{\Delta z}{1+z_{spec}})$=0.034, with no significant mean offset.  This accuracy may appear to be low relative to the $\sim$1\% reported by Ilbert et al. (2009) for their own set of photo-z's based on 30 bands of photometric data.  However, the NB210 detections which have spec-z's available in zCOSMOS are much fainter on average (the median $i^{+}_{AB}$ in a 3\arcsec\ aperture is 22 mag) than the $17.5\leq i^{+}_{AB}\leq 22.5$ sample for which the 1\% photo-z's have been derived.  Photo-z's computed by Ilbert et al. (2009) are available for 90\% of the NB210 detections with spec-z's, and when these are compared, $\sigma(\frac{\Delta z}{1+z_{spec}})$=0.041, consistent with the accuracy of the photo-z's derived here. 





\section{Composition of the Narrowband Excess Samples}\label{sec:composition}

\subsection{NB119 Excess Emitters}

\begin{figure}
\epsscale{1}
\plotone{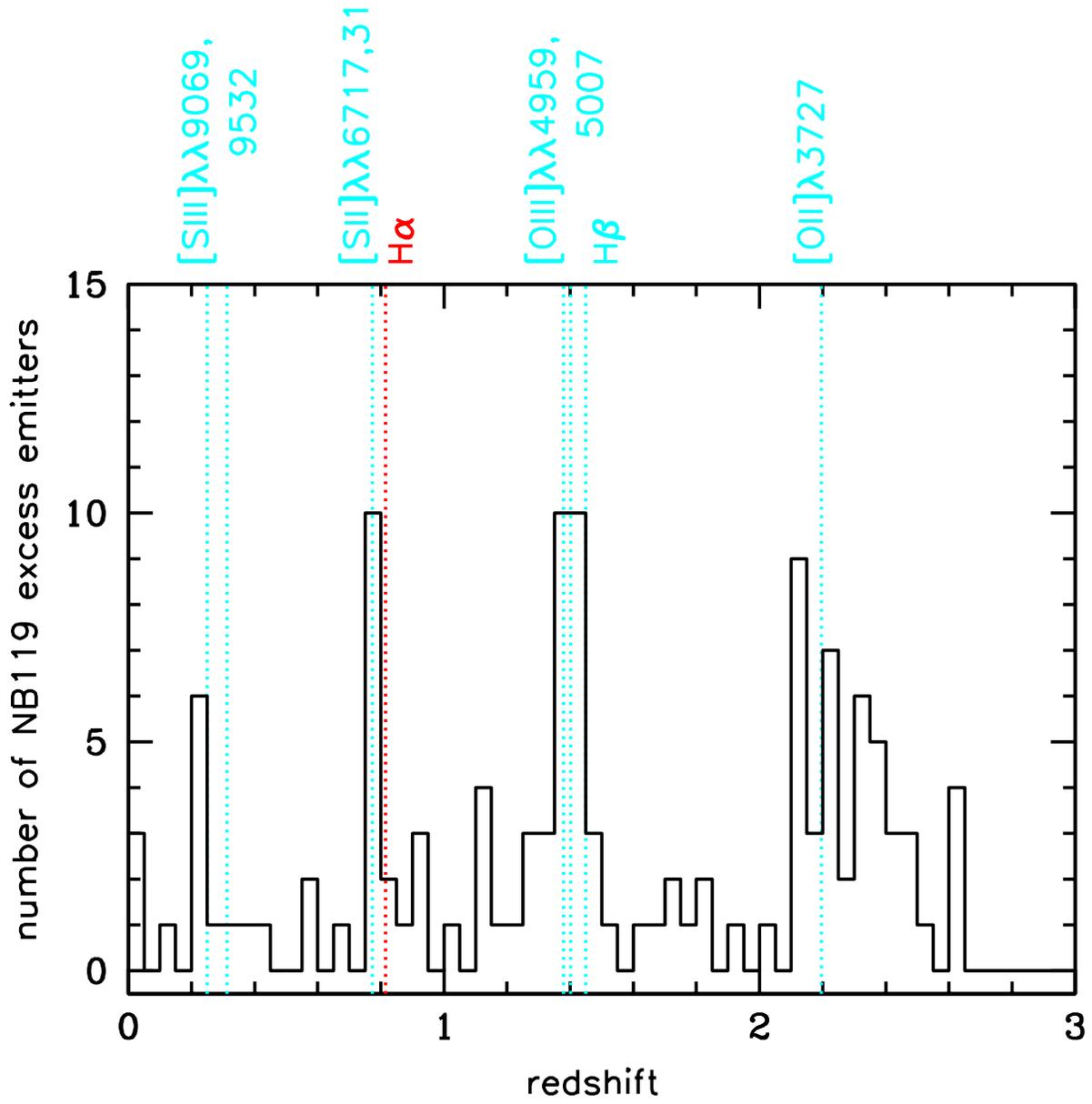}
\caption{Distribution of photometric redshifts for the NB119 excess selected sample.}
\label{fig:photz_nb118}
\end{figure}

The photometric redshifts provide a first look at the composition of the
narrowband excess samples.  As expected, the photo-z distribution shows 
a series of peaks at redshifts where strong emission-lines appear in the filter.  
Figure~\ref{fig:photz_nb118} shows that the main emission-lines detected 
in the FourStar NB119 excess sample (N=\numNBJ) are:

$\bullet$ [SIII]$\lambda\lambda$9069,9532 at $z=0.31$ and 0.24,

$\bullet$ [SII]$\lambda\lambda$6717,6731 at $z=0.76$,

$\bullet$ H$\alpha$ at $z=0.80$,

$\bullet$ [OIII]$\lambda\lambda$4959,5007 at $z=1.39$ and 1.36,

$\bullet$ H$\beta$ at $z=1.44$, and

$\bullet$ [OII]$\lambda$3727 at $z=2.19$.

The majority of the sample is at $z>1$. 
Roughly a quarter are [OIII] or H$\beta$ emitters,
while a third are [OII].  The distribution is considerably different from, but
highly complementary to 
those of previous narrowband surveys at 1.19 $\mu$m.  Past studies
reached half of the depth of the current observations or less, but
covered larger fields, and hence were dominated by foreground H$\alpha$
emitters at $z\sim0.8$ (e.g., Villar et al. 2008; Sobral et al. 2009;  Ly et al. 2011).
Here, H$\alpha$ or [SII] candidates comprise only about 10\% of the 
sample ($N\sim15$), 
but nearly all of these objects have luminosities lower 
than $10^{41}$ ergs s$^{-1}$, consistent within the expectations based 
on extrapolations of published H$\alpha$ luminosity functions at $z=0.8$,
and the volume covered in this one FourStar pointing.  
An H$\alpha$ luminosity of $10^{41}$ ergs s$^{-1}$ corresponds to 0.8 \msunyr, 
and the candidates span SFRs down to $\sim$0.3 \msunyr, 
similar to the present activity of the Large Magellanic Cloud 
(Whitney et al. 2008; Harris \& Zaritsky 2009).
Ultimately, the NewH$\alpha$ FourStar NB119 data will allow us to derive
the most stringent constraints on the faint-end slope of 
the H$\alpha$ luminosity function, and probe the star formation properties of
true dwarf galaxies at $z=0.8$.  

In principle, detection of Ly$\alpha$ at $z=8.8$ with the NB119 filter
is also possible, but unlikely given
the expected number densities of Ly$\alpha$ emitting galaxies at early times, 
and the volume and depth probed by our data (cf., Nilsson et al. 2007; 
Sobral et al. 2009).  If the Ly$\alpha$ luminosity function at $z=6.6$
(Ouchi et al. 2010) is assumed to hold at $z=8.8$, 0.4 Ly$\alpha$ emitters 
would be expected in one FourStar NB119 pointing with depth of 10$^{-17}$ 
ergs s$^{-1}$ cm$^{-2}$. However, this assumption is likely to lead to
an overly optimistic estimate as Ono et al. (2012) has reported that
the Ly$\alpha$ luminosity function appears to decrease by a factor of
two to three from $z\sim$ 6 to 7.
We are currently investigating whether any of the
NB119 sources are Ly$\alpha$ candidates, and results will be reported
in a forthcoming paper.  
Further discussion of the NB119 excess selected sample is deferred to future
work.  The remainder of this paper will focus on the NB210 excess sample, 
and dual NB210/NB119 excess emitters. 

\subsection{NB210 and Dual Narrowband Excess Emitters}

\begin{figure}
\epsscale{1}
\plotone{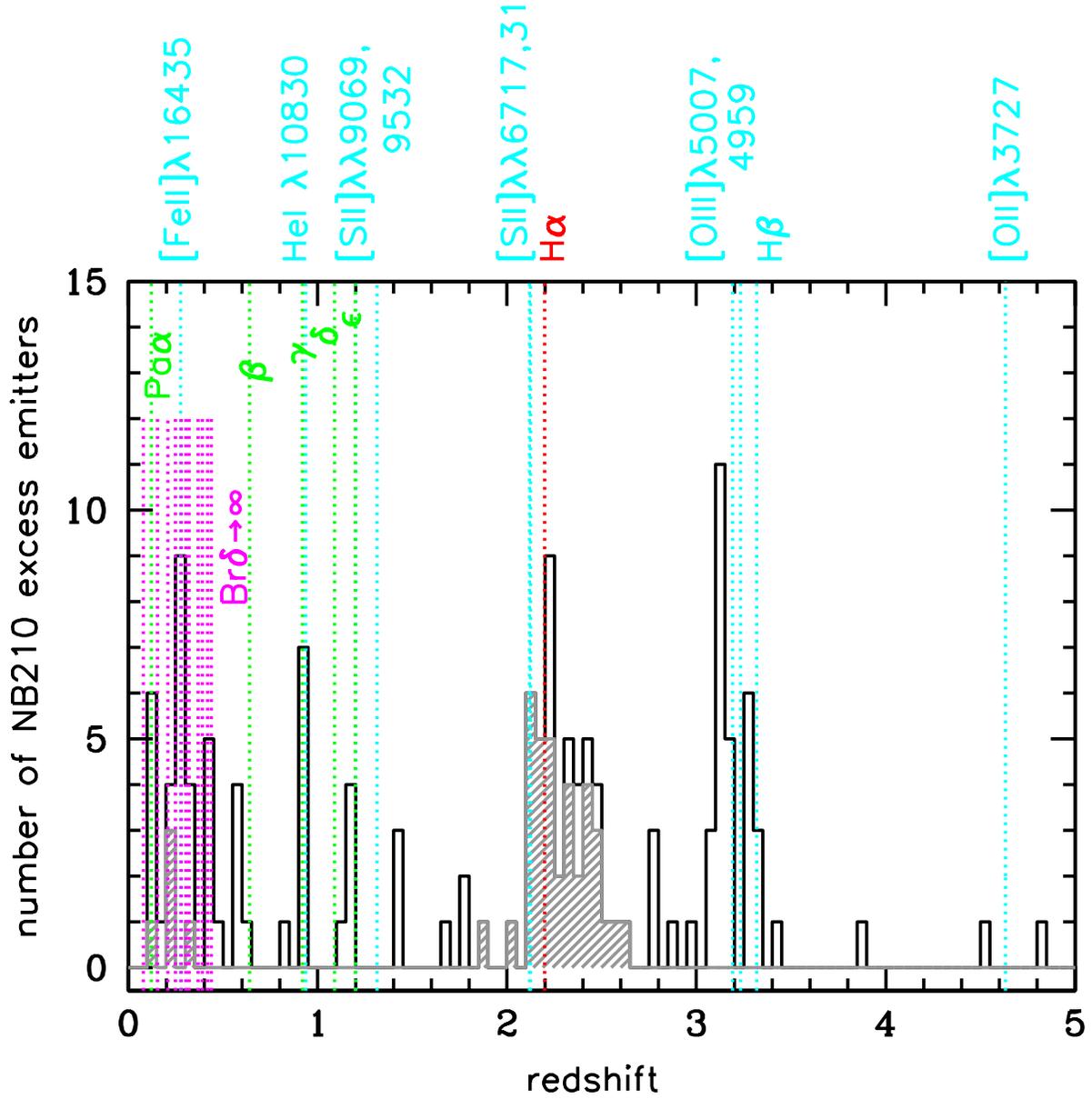}
\caption{Distribution of photometric redshifts for the NB210 excess selected sample.
Dual NB210/NB119 excess emitters are shaded.}
\label{fig:photz_nb209}
\end{figure}

In comparison, the NB210 filter is sensitive to a broader array of emission-lines.
The filter will not only detect lines redward of 1.19 $\mu$m
(such as Pa$\alpha$, Pa$\beta$ and all the high order Brackett
lines in between), but also those in the
near-IR below 1.19 $\mu$m (such as HeI$\lambda$10830 and 
Pa$\gamma$) since the NB210 filter will capture these lines at
higher redshift and thus probe significantly larger volumes relative
to the NB119 filter.
As shown in Figure~\ref{fig:photz_nb209}, the main emission-lines detected in the NB210 excess sample (N=\numNBK) are:

$\bullet$ Brackett lines from Br$\delta$ to the series limit from $z=0.08$ to 0.43, 

$\bullet$ Pa$\alpha$ to Pa$\epsilon$ at $z=0.12, 0.64, 0.92, 1.09, 1.20$ (series limit at $z=1.56$), 

$\bullet$ [FeII]$\lambda$16435 at $z=0.28$

$\bullet$ HeI $\lambda$10830 at $z=0.94$,

$\bullet$ [SIII]$\lambda\lambda$9069,9532 at $z=1.31$ and 1.20,

$\bullet$ [SII]$\lambda\lambda$6717,6731 at $z=2.12$,

$\bullet$ H$\alpha$ at $z=2.19$,

$\bullet$ [OIII]$\lambda\lambda$4959,5007 at $z=3.23$ and 3.19,

$\bullet$ H$\beta$ at $z=3.31$, and

$\bullet$ [OII]$\lambda$3727 at $z=4.62$,

The majority of NB210 excess emitters are also at $z>1$.  
Over a third are H$\alpha$ or [SII], about a quarter are
[OIII] or H$\beta$, and there may possibly be a few [OII] emitters
in the sample.  Approximately 40\% are due to H or He recombination
emission from galaxies in the foreground.  There is also a spike 
at $z\sim$0.3 containing about 15 objects,
where the shock excited [Fe II] line is expected to
appear in the NB210 filter. 
As mentioned above, only four objects
have spectroscopic redshifts from the zCOSMOS survey.  These show 
that Pa$\alpha$ at $z=0.12$ is responsible for the NB210 excess in two galaxies, 
and that the excess is from HeI$\lambda$10830 at $z=0.94$ 
in the other two.


The photo-z distribution of NB210 excess emitters that 
also have an excess in the NB119 band is shown with the shaded histogram
in Figure~\ref{fig:photz_nb209}.  About 90\% (36/41) of the 
dual narrowband excess emitters have photo-z's between 1.8 and 2.6, 
demonstrating that the technique succeeds in identifying the $z=2.2$ H$\alpha$
emitters in the sample.  Five dual excess emitters have photo-z's less
than 0.4, but examination of their SEDs and photo-z
probability distribution functions, shows that they can also be
reasonably fit with models at $z=2.2$.

\begin{figure}
\epsscale{0.8}
\plotone{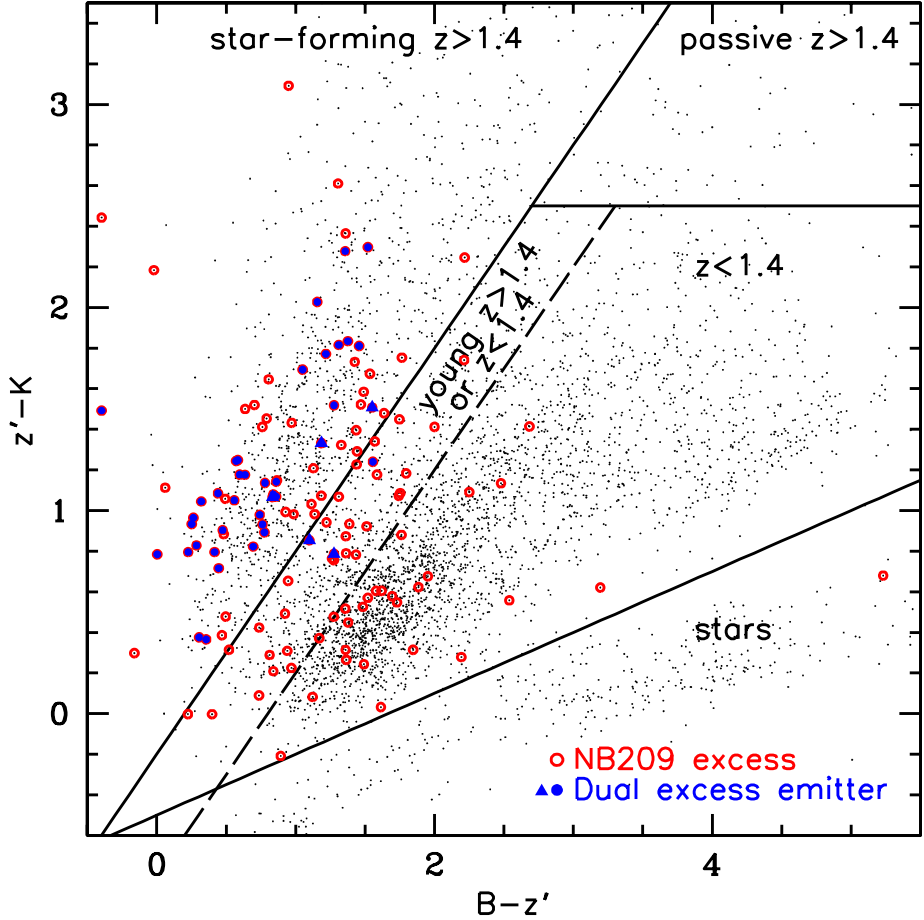}
\caption{A $BzK$ color-color diagram for all objects detected in the NB210 imaging.  NB210 excess selected objects and dual NB210/NB119 excess emitters are marked as indicated in the figure.  Criteria used to distinguish different populations of galaxies as given by Daddi et al. (2004) are also shown.  The dual excess emitters nearly all satisfy the criteria for classification as ``$sBzK$" galaxies, indicating that they are star-forming galaxies at redshifts between 1.4 and 2.5, and demonstrating that the dual excess selection technique succeeds in identifying the $z=2.2$ H$\alpha$
emitters in the sample. Five dual excess emitters have photo-z's less than 0.4 (blue triangles), but their SEDs can also be reasonably fit with models at $z=2.2$, as corroborated by their BzK colors.}
\label{fig:BzK}
\end{figure}

The effectiveness of the NB210/NB119 dual narrowband excess technique
for selecting a clean sample of H$\alpha$ at $z=2.2$ can also be shown
by plotting the objects on a {\it BzK} color-color diagram.  The {\it BzK} diagram
has frequently been used to obtain samples of galaxies with redshifts between about 1.4 and 2.5 with relatively high completeness and low contamination, and also to distinguish between star-forming and passively evolving galaxies, as well as stars (e.g., Daddi et al. 2004; Kong et al. 2006; McCracken et al. 2010).  The method was introduced, calibrated and tested by Daddi et al. (2004) using both spectroscopic samples and stellar population synthesis models.  In Figure~\ref{fig:BzK}, the original criteria given by Daddi et al. (2004) are marked with solid lines.\footnote{The empirically derived transformation given by McCracken et al. (2010) to account for differences between the Subaru $B_J$ filter used to image the COSMOS field, and the $B$-VLT filter used by Daddi et al. (2004) is applied to the data shown.}  All but three of the dual excess emitters (blue solid symbols) satisfy the criteria for classification as intermediate redshift star-forming galaxies.  Even the three objects that fail to meet the criteria fall in a region slightly below the star-forming {\it BzK} boundary, $(z-K)-(B-z) > -0.2$, in which populations dominated by very young populations ($\lesssim10^8$ yrs) can exist.  Daddi et al. (2004) notes that to capture such galaxies, the boundary must be extended to $(z-K)-(B-z) > -0.8$ (dashed line), but that this comes at the cost of higher contamination by populations at $z<1.4$.  The five dual emitters that have photo-z's less than 0.4 (triangles) straddle the formal $(z-K)-(B-z) > -0.2$ star-forming {\it BzK} boundary, with two falling outside the boundary.  Optical spectroscopy can be performed to check whether these are in fact low redshift galaxies.  In any case, the contamination of the dual excess sample with non-H$\alpha$ emitters will be very low, $\lesssim 5\%$. 

Figure~\ref{fig:photz_nb209} also shows that there are NB210 excess objects that have photo-z's which would make them H$\alpha$ candidates, but do not have NB119 excess detections.  If the photo-z range for H$\alpha$ is taken to be between 1.8 and 2.6 (as defined by the range of the majority of dual excess emitters), there are nine such objects.  The two most likely explanations for these objects is that the NB210 excess is due to [SII]$\lambda\lambda6717,31$, or that the depth of the NB119 image was not sufficient to detect [OII].  

\begin{figure}
\epsscale{0.5}
\plotone{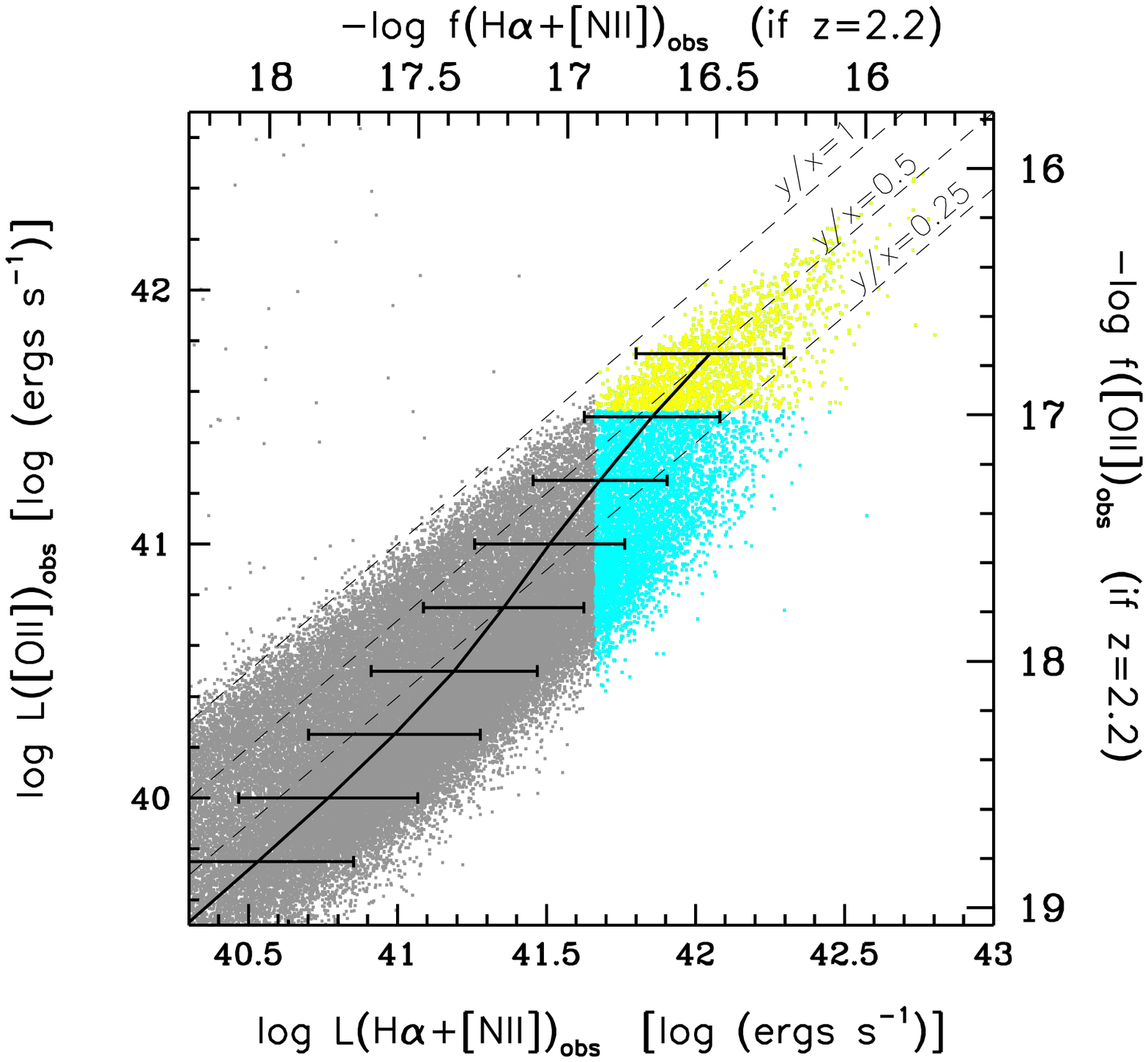}
\plotone{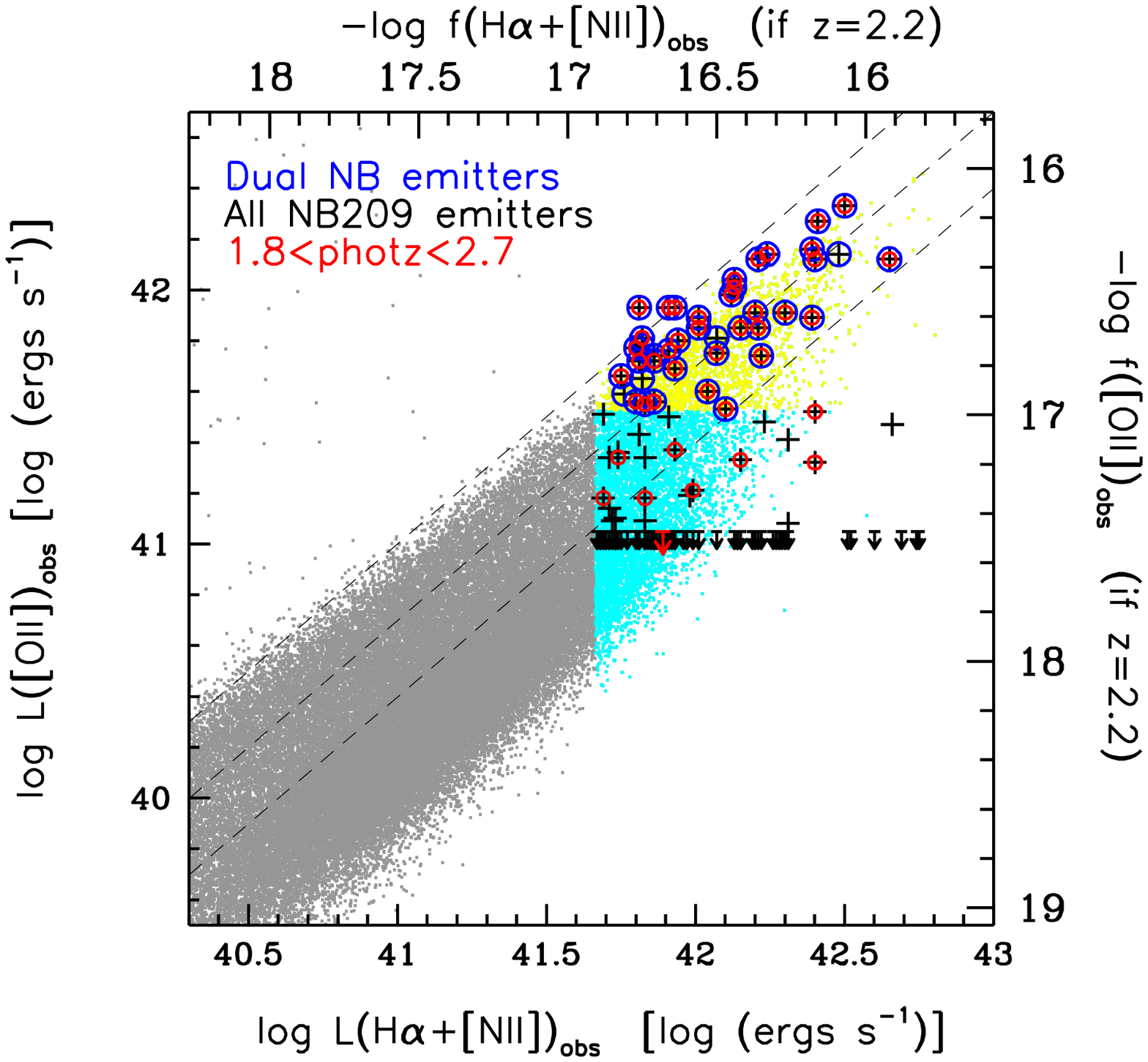}
\caption{Top: [OII]$\lambda$3727 and H$\alpha$ luminosities for local galaxies in the SDSS DR7 sample.  The emission-line flux sensitivity limits of the NewH$\alpha$ FourStar narrowband imaging are illustrated by showing the SDSS galaxies whose H$\alpha$ emission would be detectable in the NB210 data if they were at $z=2.2$ (yellow and cyan regions), and those which would also have their [OII] detected in the NB119 data (cyan region).  Median values computed in bins of L([OII]) are plotted along with the 1$\sigma$ widths of the distributions.  Bottom: The sample of NB210 excess emitters are overplotted (black crosses), and dual NB210/NB119 excess emitters are indicated (blue open circles). Objects which have photo-z's which would make them H$\alpha$ candidates are also marked (red open circles). The dual narrowband excess technique, as implemented here, is estimated to identify $\gtrsim$80\% of $z=2.2$ H$\alpha$ emitters in the sample. }
\label{fig:oiiha}
\end{figure}

To examine the latter possibility, the H$\alpha$+[NII] and [OII] fluxes 
for the objects in the NB210 excess sample are plotted in 
Figure~\ref{fig:oiiha}, and compared with those observed 
locally from the Sloan Digital Sky Survey (SDSS) Data Release 7 
sample (DR7; Abazajian et al. 2009).  Emission-line fluxes are
computed from the FourStar narrowband and broadband photometry in the usual
way (e.g., see equations 9 and 10 in Ly et al. 2011).
Spectroscopic emission-line fluxes for the SDSS sample are made publicly
available by the MPA/JHU group\footnote{http://www.mpa-garching.mpg.de/SDSS/DR7/}.
The fluxes have been extracted from Gaussian fits to the line profiles, 
where the continuum has been subtracted using a best-fit model generated 
by a combination of simple stellar population at various ages from 
the Bruzual \& Charlot spectral synthesis code (Bruzual \& Charlot 2003). 
Corrections for Galactic foreground reddening are applied.      
SDSS spectra for a total of 818,333 galaxies at $z<0.7$ are available, but  
here we only consider the sub-sample of $\sim$60,000 star-forming galaxies with 
$0.005 < z < 0.22$ (following Brinchmann et al. 2004) and 
an empirically estimated fiber aperture correction less than 2.  
The 3$\sigma$ sensitivities of the narrowband 
imaging to these lines is also illustrated by indicating which of the 
SDSS galaxies would be detectable if they were instead at $z=2.2$.  Clearly, 
{\it if} the full range of local galaxies exists at redshift of two, current
near-IR narrowband surveys, of which NewH$\alpha$ is one of the most 
sensitive to faint line emission, would still only be able to probe
the most luminous H$\alpha$ and [OII] emitters.
 

In Figure~\ref{fig:oiiha}, all of the dual narrowband excess emitters lie in the region above both the NB119 and NB210 3$\sigma$ line flux sensitivity limits as expected (yellow region).  The region where objects would be only detected with a NB210 excess is indicated in cyan.  Here, more uncertain NB119 excess fluxes (which would correspond to [OII] if the objects were indeed at $z=2.2$) down to the 1$\sigma$ limit (marked by the row of upper-limit symbols) are also plotted.  It is clear that even without additional photo-z or SED information, some objects can be eliminated as H$\alpha$ candidates because they would have [OII]/H$\alpha$ ratios that are unphysical.  However, the nine galaxies that do not have 3$\sigma$ excess detections in NB119, but yet have photo-z's which would make them H$\alpha$ candidates, all would have [OII]/H$\alpha$ ratios that fall within the range of physically plausible values.  Further, all except one have nominal NB119 excess detections above 1$\sigma$, which would not be expected if the NB210 excesses were mostly due to [SII].  This suggests that the majority of these objects are H$\alpha$ emitters at $z=2.2$.  Therefore, the dual narrowband technique, as implemented here, is estimated to identify $\gtrsim$80\% of $z=2.2$ H$\alpha$ emitters detected in the NB210 excess sample.  Near-IR spectroscopy will be required to confirm this success rate.

\section{Constraints on the H$\alpha$ Luminosity Function at $z=2.2$}

With a clean, complete sample of H$\alpha$ emitters in hand, constraints on the $z=2.2$ H$\alpha$ luminosity function can be computed.  We follow the procedures discussed in detail in Ly et al. (2011), which presents our results on the H$\alpha$ luminosity function at $z=0.8$, using NB119 imaging taken with the much wider FoV NEWFIRM camera on the KPNO 4m telescope.  Here, we outline the procedure, pointing out differences between the analyses when they arise.

The key steps include (1) transforming the narrowband photometric excesses into dust-corrected H$\alpha$ luminosities; (2) computing the effective volume of the observations; (3) estimating and applying completeness corrections; and (4) fitting a function, usually that of Schechter (1976) (but see Salim et al. 2012, in preparation), to model the distribution.  The step where methods which have been previously applied for analogous studies at lower redshifts cannot be directly transplanted for the current analysis is the first one.  

To compute the intrinsic H$\alpha$ luminosity from the narrowband excess, two main corrections are required, one for emission from [NII]$\lambda\lambda$6548,83 (since the flux from these lines will contribute to the narrowband excess as R$\sim$100 filters are generally used), and another for dust attenuation within the galaxy.  Because measurements of [NII]/H$\alpha$ and A(H$\alpha$)\footnote{attenuation of H$\alpha$ emission in magnitude units} are not immediately available for the individual members of a sample, coarse empirical scaling relationships calibrated on local samples such as the SDSS have been employed to estimate their values (e.g., Villar et al. 2008; Sobral et al. 2009; Lee et al. 2009; Ly et al. 2011).  For example, in previous H$\alpha$ narrowband studies, relations providing [NII]$\lambda$6584/H$\alpha$ as a function of the rest-frame H$\alpha$ equivalent width (Villar et al. 2008), and the intrinsic H$\alpha$ luminosity as a function of the observed H$\alpha$ luminosity (Hopkins et al. 2001) have been used.  However, it is well known that such relationships have scatters of a factor of two or more (e.g., Villar et al. 2008; Kennicutt et al. 2008; Lee et al. 2009; Momcheva et al. 2012).  Moreover, it is unclear whether these scaling relations will be valid at $z=2$, as the gas metallicities and dust reddenings of galaxies of a given stellar mass are known to decrease with increasing redshift (Erb et al. 2006a; Reddy et al. 2006; Maiolino et al. 2008; Reddy 2010).

Though we do not yet have independent measurements of [NII]/H$\alpha$ and A(H$\alpha$) for our $z=2.2$ H$\alpha$ sample, we can take a first look at where these values are likely to lie via comparison to the well-studied rest-frame UV-selected $z\sim$2 sample of Erb et al. (2006a,b,c), for which  H$\alpha$ spectroscopy has been obtained, and dust reddenings have been computed both from SED fitting and the UV continuum slope (Erb et al. 2006c).  Such an exercise also serves to place the current dataset in the more familiar context of a broadly referenced sample of star-forming galaxies at $z\sim$2.

\begin{figure}
\epsscale{1.1}
\plottwo{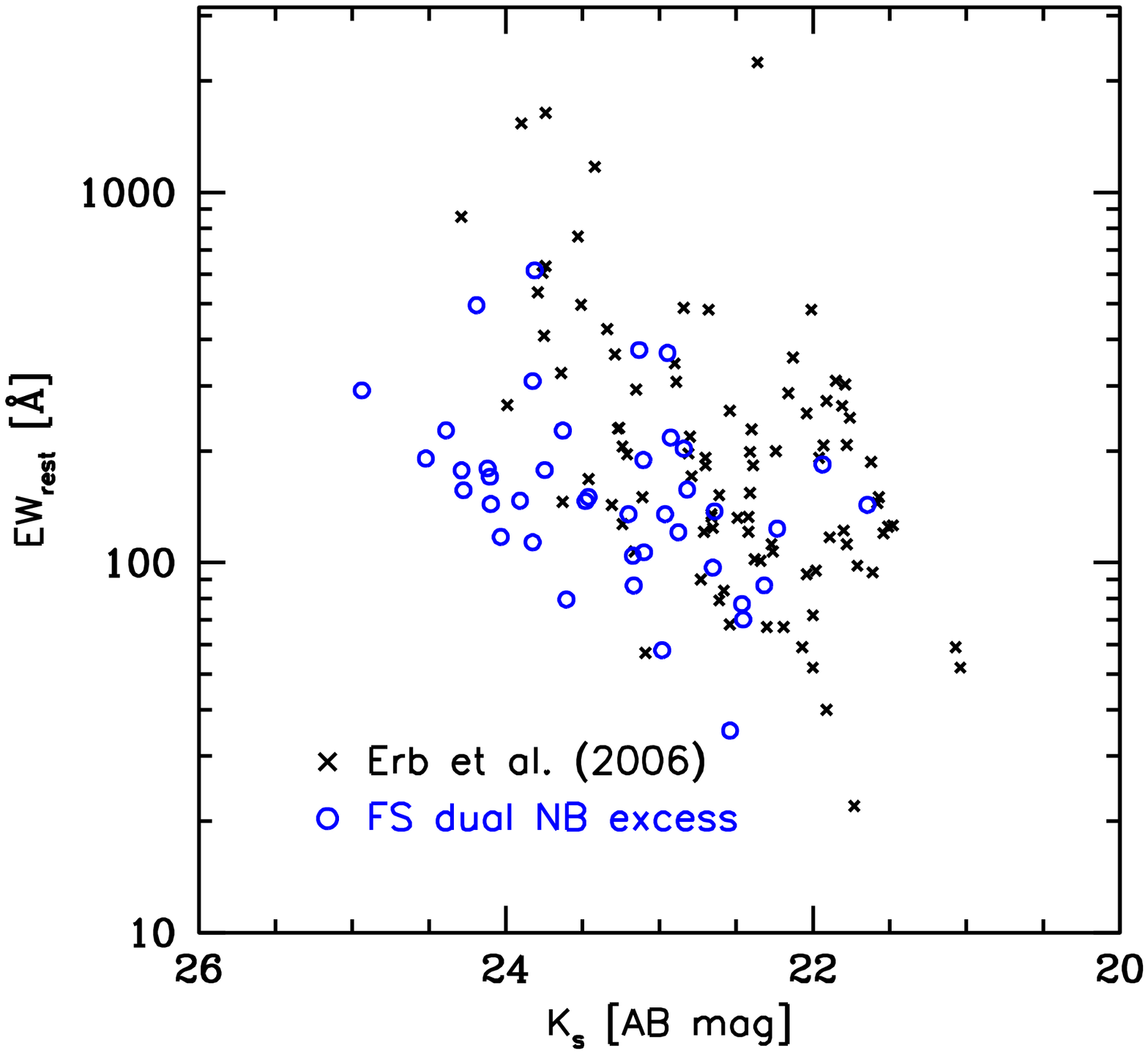}{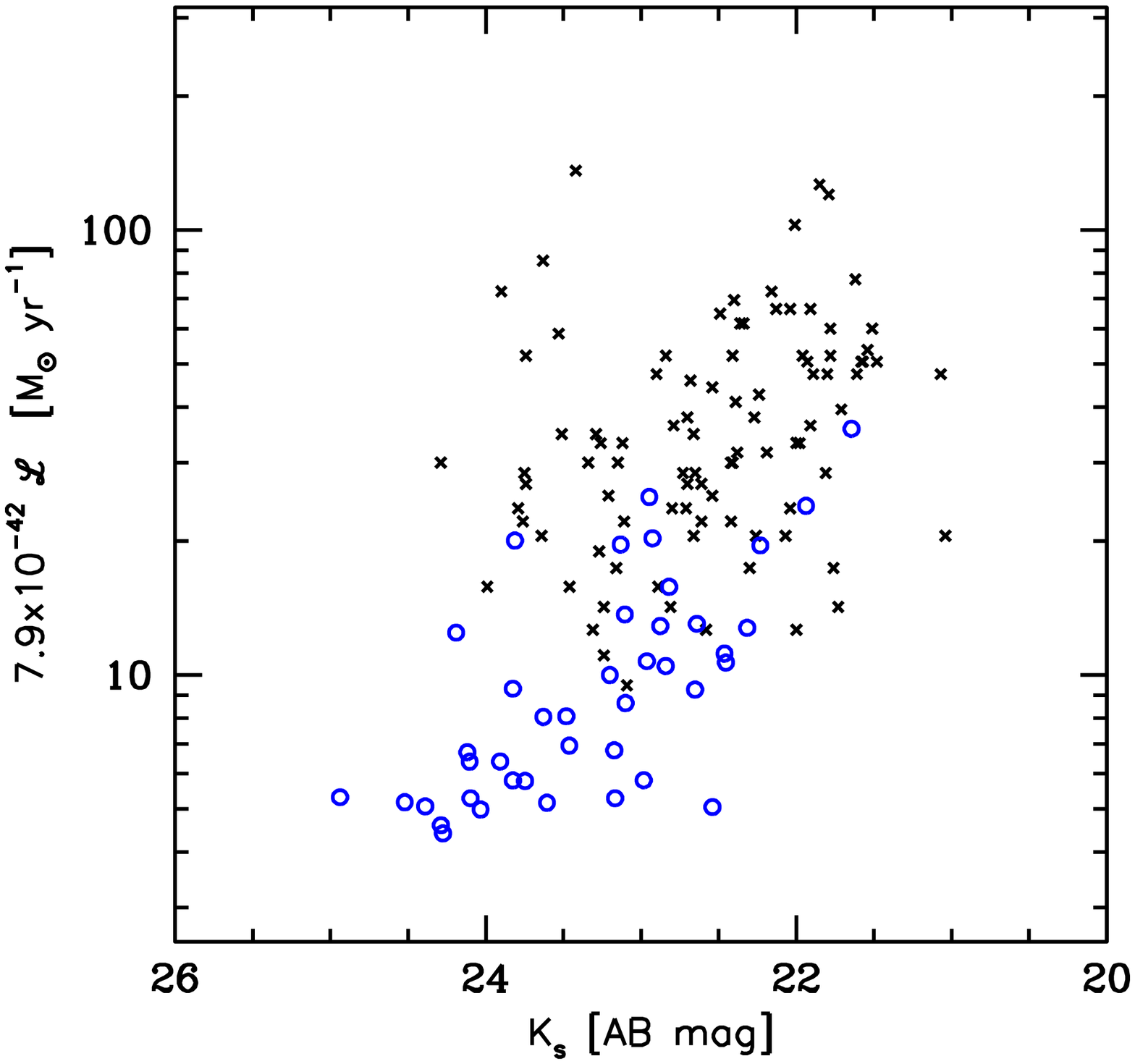}
\caption{Comparison of the $z=2.2$ dual narrowband selected H$\alpha$ emitters (blue circles) with the rest-UV selected $z\sim2$ sample of Erb et al. (2006c) (black crosses).  In both panels the plotted measurements for the H$\alpha$ emitters include both H$\alpha$ and [NII]$\lambda\lambda$6548,83, while those for the UV-selected sample only include H$\alpha$.  A factor of two correction for spectroscopic slit losses has been accounted for in the line measurements of the UV selected sample, as determined by Erb et al. (2006c).  All quantities shown are as-observed, with no correction applied for dust attenuation.  The left panel shows the rest-frame equivalent width, while the right panel shows the line luminosity, scaled by 7.9$\times10^{-42}$ to give indicative star formation rates (Kennicutt 1998).  Both are plotted as a function of the $K_s$ AB magnitudes.  On average, narrowband H$\alpha$ sample extends the luminosities and SFRs probed for $z\sim$2 star-forming galaxies lower by a factor of at least two.}
\label{fig:erb}
\end{figure}

In Figure~\ref{fig:erb}, the rest-frame equivalent width, and the line luminosity, scaled by 7.9$\times10^{-42}$ to give approximate star formation rates (Kennicutt 1998), are plotted against the $K_s$ magnitude for both samples.  For the UV-selected sample, the measurements are made from slit spectroscopy, and so are for only the H$\alpha$ line.  
These measurements include a factor of two correction to account for slit losses, as determined by Erb et al. (2006c).  For the FourStar H$\alpha$ emitters, the plotted values include contributions from both [NII] and H$\alpha$.  No dust corrections have been applied to either sample.  $K_s$ magnitudes on the AB scale are shown for both, where $K_s$ $[$AB$]$=$K_s$ $[$Vega$]+$1.83 is applied to the Erb data.  
 
Although the EW range (and hence range of specific SFR) spanned by the two samples are similar, it is clear that the H$\alpha$ emitters probe SFRs and continuum luminosities (and hence stellar masses) that are significantly lower than the UV selected sample.  The medians of the distributions for both quantities are at least a factor of two smaller for the H$\alpha$ emitters.  Hence, if the [NII]/H$\alpha$ values and reddenings derived for the UV-selected sample were to be used to formulate corrections for the H$\alpha$ emitters, it seems that some extrapolation of the corrections would be required.

\begin{figure}
\epsscale{1.1}
\plottwo{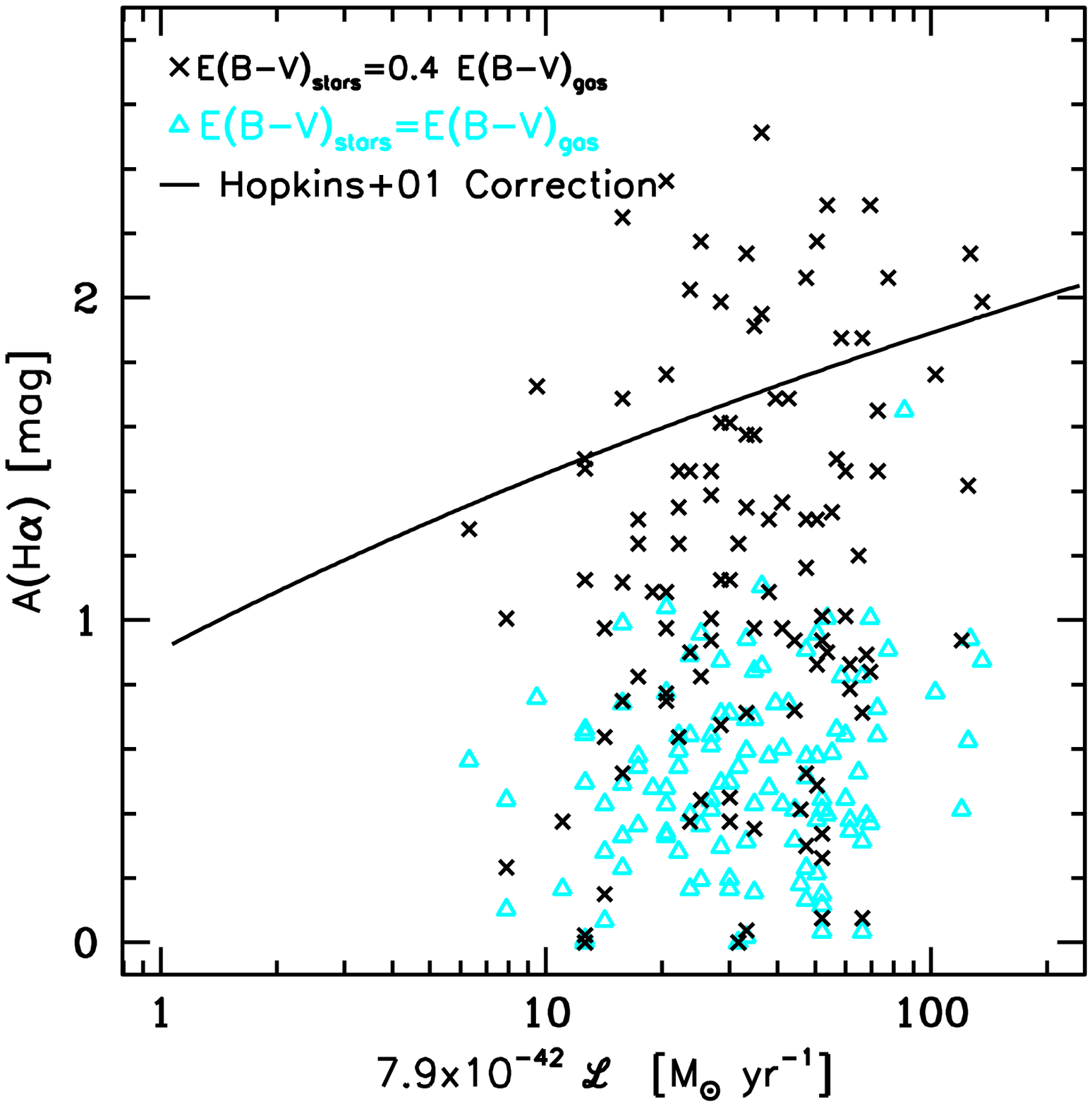}{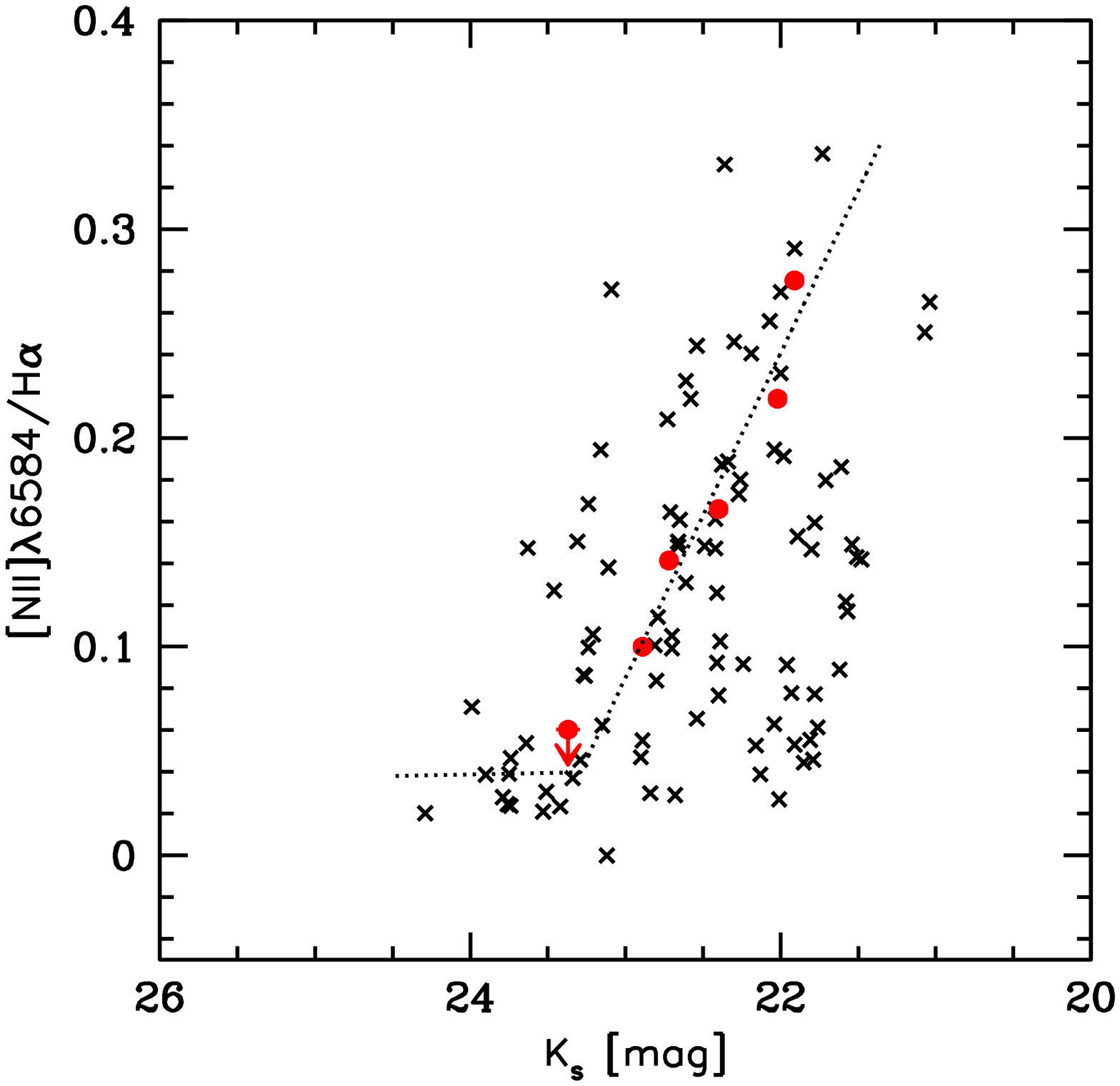}
\caption{Tests of local empirical relations, which are frequently used to correct for dust reddening and the contribution of [NII], using the measurements of Erb et al. (2006a,b,c).  Left: The attenuation of H$\alpha$ for the Erb UV selected sample is plotted as a function of observed SFR, and the curve shows the relation given by Hopkins et al. (2001).  Erb et al. (2006b) provide E(B-V)'s derived by SED fitting, and these are scaled using the Calzetti et al. (2000) extinction law to give the attenuation in H$\alpha$, assuming that reddening in the gas is about twice that of the stars (black crosses) and that it the same as that of the stars (light blue triangles).  In either case, the Hopkins prescription will over-predict the attenuation.  Right: EW(H$\alpha$+[NII]$\lambda$6583) is used to predict [NII]$\lambda$6584/H$\alpha$ for the Erb UV selected sample using the local relation of Villar et al. (2008). The results are plotted as a function of $K_s$ magnitude, to enable comparison with the Erb et al. (2006a) stacked spectroscopic [NII]/H$\alpha$ measurements (red circles). Here the local relation yields reasonable predictions, albeit with large scatter.  To correct for the contribution of [NII] to the narrowband fluxes for the $z=2.2$ H$\alpha$ emitters, the relation shown by the dotted line is used to estimate the [NII]$\lambda$6584/H$\alpha$ ratio from the $K_s$ magnitude.}
\label{fig:erb_corr}
\end{figure}

We first examine the reddenings for the UV-selected sample found by Erb et al. (2006b) in the left panel of Figure~\ref{fig:erb_corr}.  Here, 
A(H$\alpha$) is plotted as a function of the observed SFR, to enable comparison with the locally calibrated Hopkins et al. (2001) relation.  Erb et al. (2006b) provide E(B-V)'s derived by SED fitting, and these are scaled using the Calzetti et al. (2000) extinction law to give the attenuation in H$\alpha$.  A(H$\alpha$) is computed both by assuming that the reddening in the gas is about twice that of the stars, which is found locally (Calzetti 2000); and, that it the same as that of the stars, which seems to be supported by intermediate redshift observational studies (Erb et al. 2006b, but see Forster-Schreiber et al. 2009).  There is no correlation with the observed SFR, and in both cases, the Hopkins prescription over-predicts the attenuation.  The median H$\alpha$ attenuation is 0.5 mag when E(B-V)$_{stars}$=E(B-V)$_{gas}$ is assumed, and so for the current analysis, we simply adopt this fixed value to perform the dust correction.

Next, in the right panel of Figure~\ref{fig:erb_corr}, the equivalent widths from Erb et al. (2006c) are used used to predict [NII]$\lambda$6584/H$\alpha$ using the local relation of Villar et al. (2008).  The estimates are plotted as a function of $K_s$ magnitude.  This enables comparison with the Erb et al. (2006a) measurements, which were performed on long-slit spectra, stacked in bins of stellar mass.  $K_s$ magnitudes are provided for each mass bin.   Figure~\ref{fig:erb_corr} shows that the local relation actually yields reasonable predictions, albeit with large scatter.  The consistency between the locally-based estimates and the $z\sim2$ measurements may initially seem surprising given the known evolution of the mass-metallicity relation, but is likely a consequence of the ``Fundamental Metallicity Relation,'' or FMR, proposed by Mannucci et al (2010).  The FMR is a more general relation between mass, SFR and metallicity, with a small residual dispersion in the metallicity of only $\sim$12\% based upon the SDSS, and is shown to hold for galaxies to $z\sim$2.5.  Collapse in one dimension yields a correlation between metallicity and specific SFR, which is traced by the H$\alpha$ EW.  Here, although the Villar et al. (2008) scaling relation appears to be valid, we choose to use a linear fit to the Erb et al. (2006c) measurements to estimate the  [NII]$\lambda$6584/H$\alpha$ values for the H$\alpha$ emitters from their $K_s$ magnitudes.  The fit is given by [NII]$\lambda$6584/H$\alpha=-0.15K_s+3.67$.  For $K_s>23.3$, [NII]$\lambda$6584/H$\alpha=0.04$ is assumed.  The validity of both this relation and the 0.5 mag dust correction will need to be tested in the future with follow-up near-IR spectroscopy.

Steps (2), (3), and (4) are then performed by following the methods used in Ly et al. (2011),
which we briefly review here.  
The effective volume is computed, accounting for the deviation of the filter profile 
from a perfect top-hat function. 
The maximum surveyed volume is $3.83\times10^5$ Mpc$^3$ deg$^{-2}$
for H$\alpha$ emission-line fluxes above $2.7\times10^{-17}$ erg cm$^{-2}$ s$^{-1}$, and
decreases to half of this volume at an H$\alpha$ emission-line flux of $1.35\times10^{-17}$
erg cm$^{-2}$ s$^{-1}$.  The total area covered by the stacked narrowband observations is 127 arcmin$^{-2}$, and hence the resulting maximum volume is 1.36$\times$10$^4$ Mpc$^3$.  Next, Monte Carlo simulations are
performed to estimate the completeness fraction at a given H$\alpha$ luminosity.  
The simulations are used to compute the probability of detecting a source of a given continuum brightness
and emission-line flux (i.e., NB210 excess), based upon the sensitivity of the
narrow and broad-band images.  We follow a maximum-likelihood 
approach where assumptions about the
true distributions of the $z\sim2$ star-forming population are varied to best
reproduce the observed H$\alpha$ LF and EW distribution. 
To model the intrinsic H$\alpha$ EWs, log-normal Gaussian distributions are adopted, where
the {\it rest-frame} median logarithmic EW and standard deviation 
are varied in 0.1 dex increments between 1.8 and 2.3; and 0.1 and 0.6, respectively.
The $K_s$-band luminosity function for $z\sim2$
star-forming BzK galaxies, which is determined from data published in Ly et al. (2011b),
is assumed.  
With the $K_s$-band brightnesses and H$\alpha$ EW's, 400,000 sources for each of
6$\times$6 different EW model distribution are generated.  Noise is then added to the sources, 
and the selection criteria described in Section 5.2 are applied to calculate 
the completeness as a function of H$\alpha$ flux. The EW model that best
reproduces both the observed H$\alpha$ LF and EW distribution is described by a median 
log(EW$_{rest}$) of 2.1 and $\sigma$[log(EW)] =0.3 dex.  The completeness of the sample is
99\% at a flux of 2.2 $\times$ 10$^{-17}$ ergs s$^{-1}$ cm$^{-2}$ and
50\% at  1.2 $\times$ 10$^{-17}$ ergs s$^{-1}$ cm$^{-2}$.  For comparison, the median log(EW$_{rest}$) is 1.5 for local galaxies along the star-forming sequence (Lee et al. 2007).  The characteristic specific SFR, which is traced by EW(H$\alpha$), is a factor of $\sim$3 higher at $z\sim2$ (c.f., Noeske et al. 2007).

\begin{figure}
\epsscale{1.1}
\plottwo{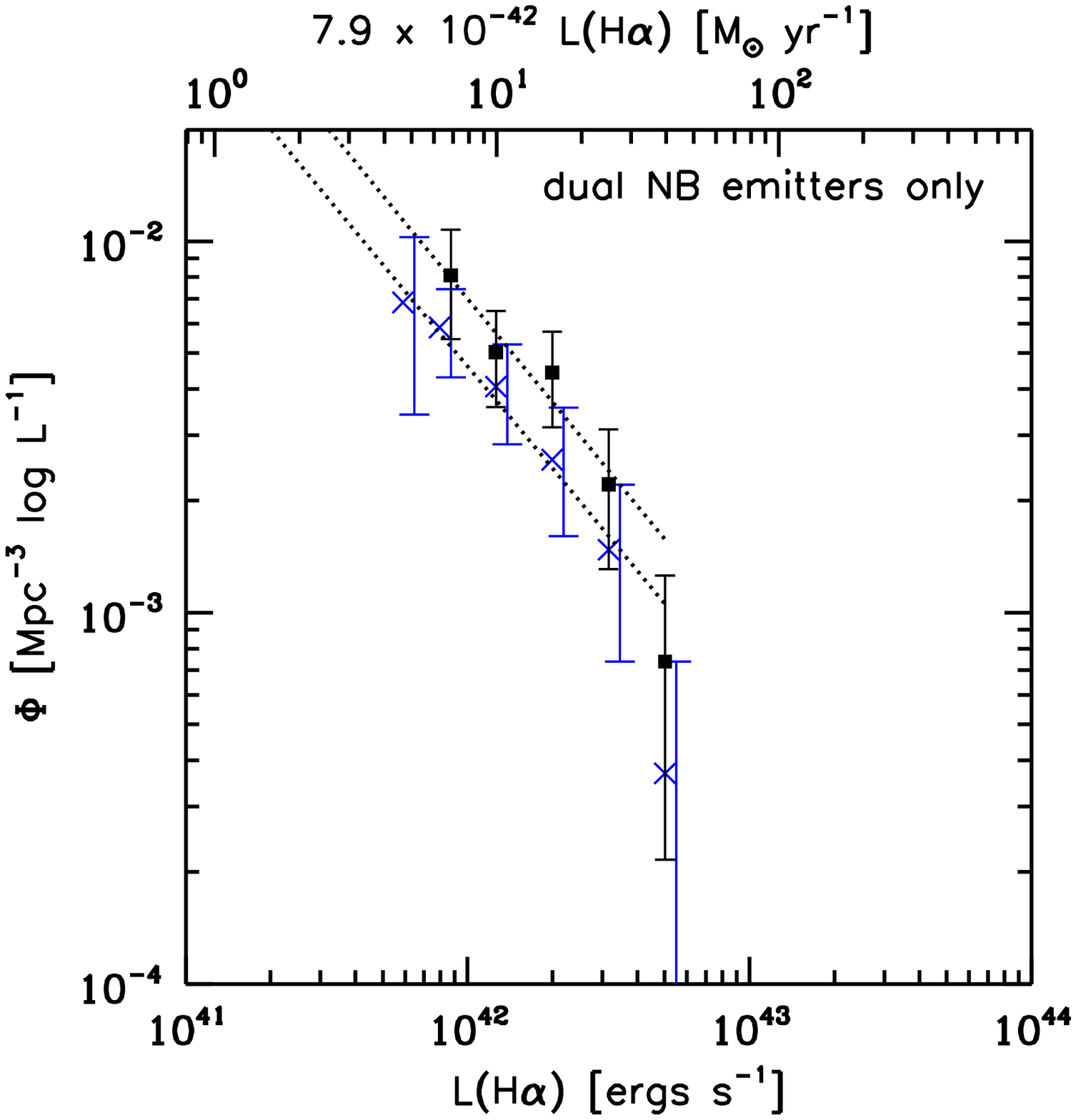}{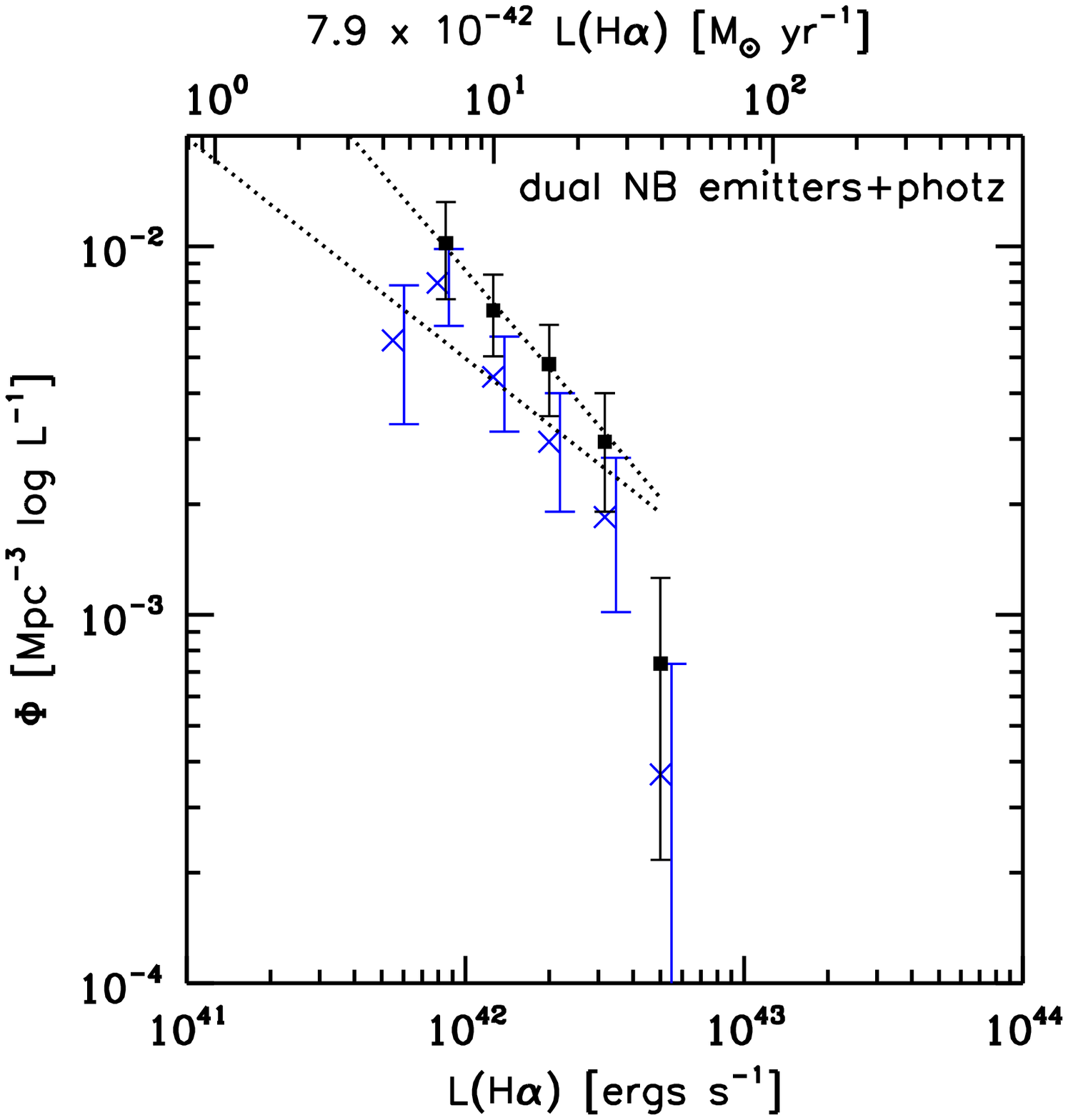}
\plottwo{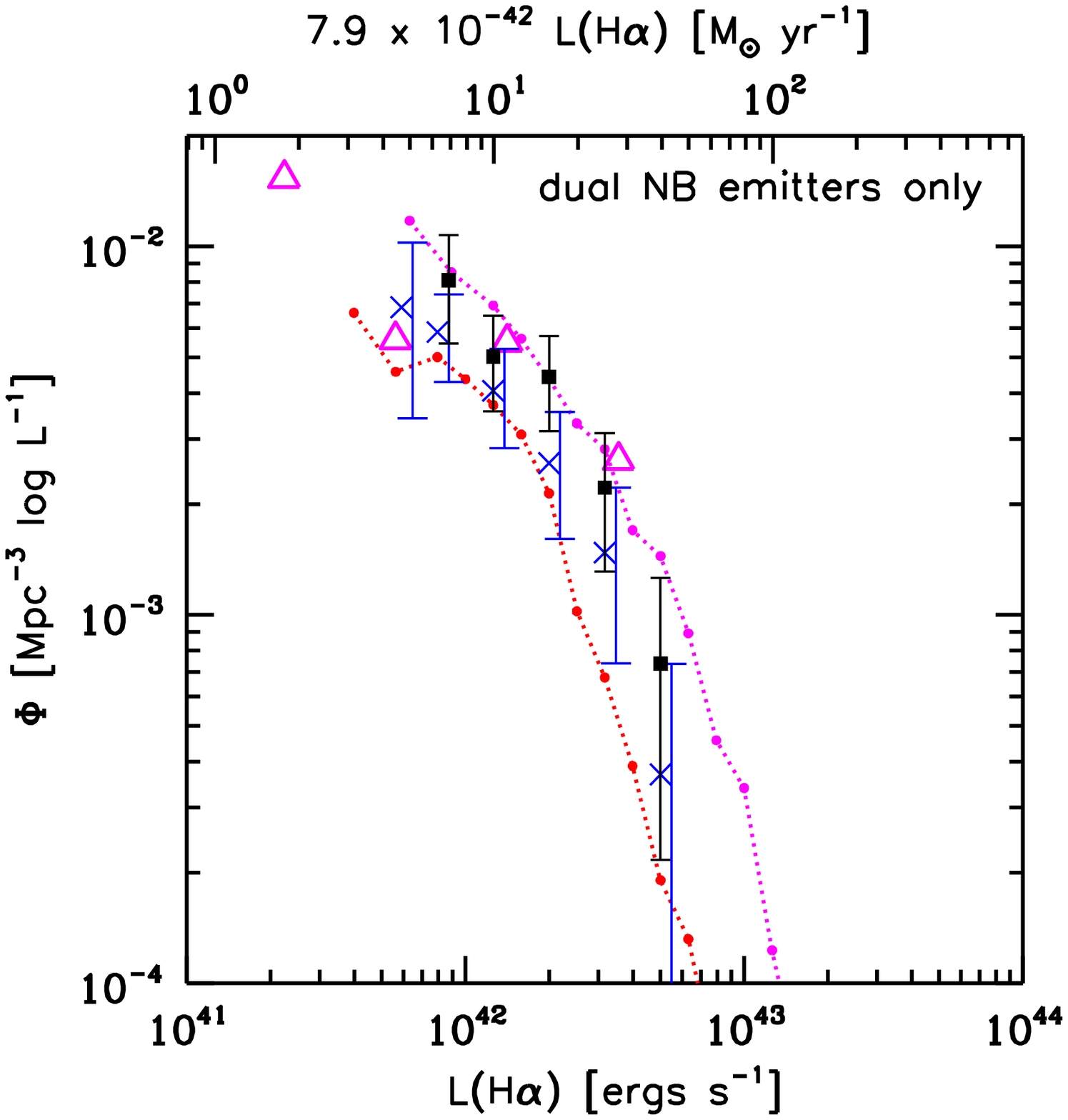}{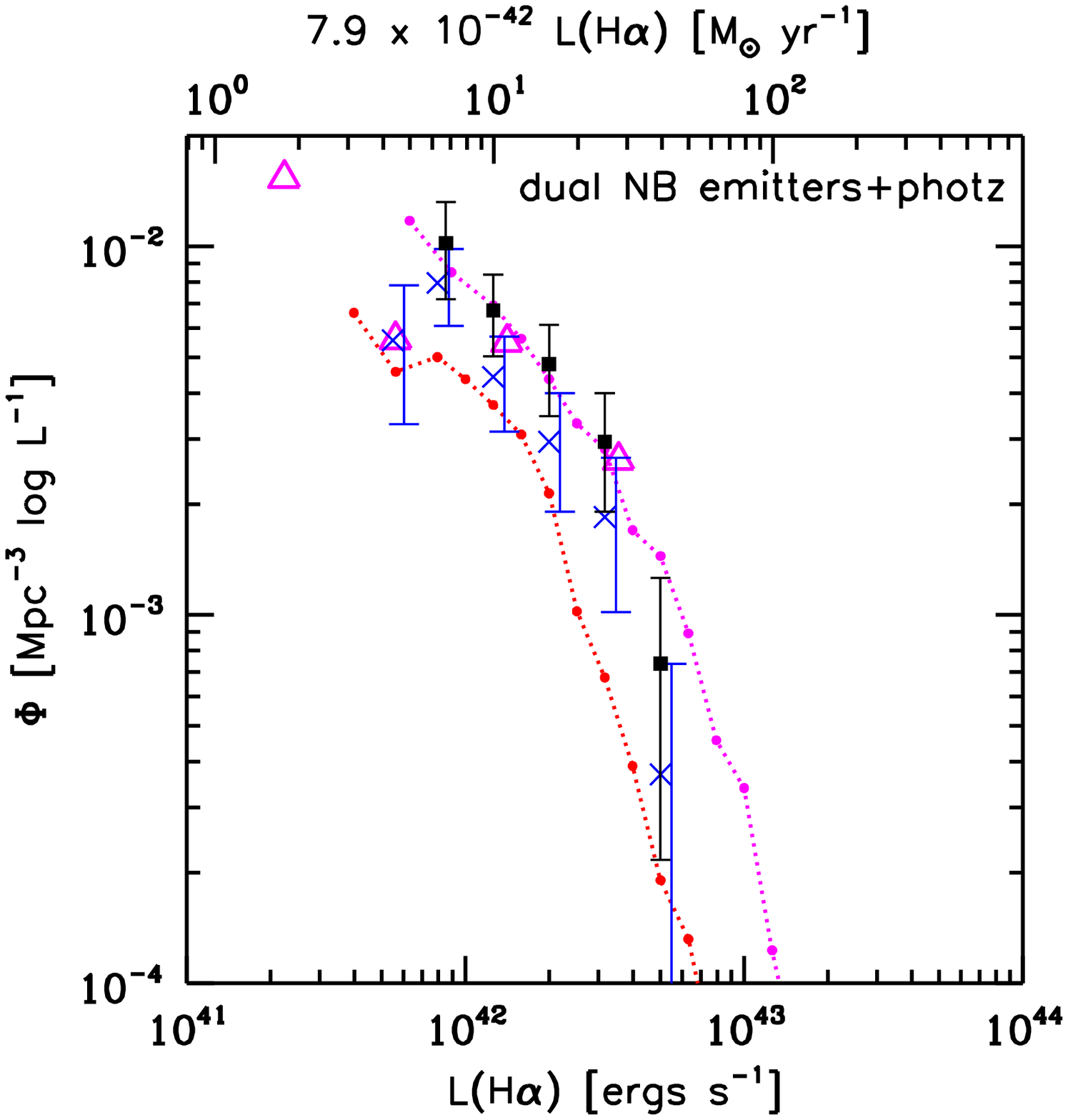}
\caption{H$\alpha$ luminosity functions at $z=2.2$.  The left panels show the LFs based on the dual narrowband excess selected sample (N=41), while the right panels show those based on the dual narrowband excess selected sample plus NB210 excess galaxies with photo-z's between 1.8 and 2.6 (N=50).  In each plot, the raw observed LF (no corrections for [NII], dust, or completeness) is indicated by the blue crosses, while the LF with all of these corrections applied is given by the black squares.  Error bars for the former are shown slightly offset in L(H$\alpha$) for clarity.  The top and bottom rows are identical, except that results from other recent H$\alpha$ narrowband studies are overlayed. LFs from Hayes et al. (2010) (magenta triangles), and from Sobral et al. (2012) (red curve: [NII] correction applied]; magenta curve: [NII], dust and completeness corrections all applied) are shown.  Dust attenuation corrections have been made consistent with those applied here. }
\label{fig:lf}
\end{figure}

Figure~\ref{fig:lf} presents the H$\alpha$ luminosity functions.   
The left panels show the LFs based on the dual narrowband excess selected 
sample (N=41), while the LFs in the right panels also include the NB210 excess galaxies 
with photo-z's between 1.8 and 2.6 (N=50). 
In every plot, there are two sets of points corresponding to (1) the raw H$\alpha$+[NII] 
luminosities, and (2) the fully-corrected luminosities, where the contribution of the 
[NII] lines, 0.5 mag of dust attenuation, and survey incompleteness have all been taken
into account.  The data points for all four luminosity functions are given in Table~\ref{tab:lf}. 
The bottom row is identical to the top row, except that the results of
Hayes et al. (2010; hereafter H10) and Sobral et al. (2012; hereafter S12) are overplotted.  

Both the raw and fully-corrected LFs appear to rise steeply at the faint end.
Single power-law fits to the faintest three, four or five bins for the four
samples shown in Figure~\ref{fig:lf} yield power law exponents in the range
$-1.60\leq\alpha\leq-1.93$, with 1$\sigma$ uncertainties
of $\sim0.3$.  These values are consistent with the
H10 VLT/HAWK-I result of $\alpha=-1.77\pm0.21$, 
which was based on a
photometric redshift selected 2.095 $\mu$m narrowband excess sample (N=55). 
Such results may imply that the star formation rate distribution function
gradually steepens from the present day value of $\alpha=-1.2\pm0.2$ (e.g. Perez-Gonzalez et al. 2003), as has been suggested previously 
(e.g., Hopkins et al. 2000; Reddy \& Steidel
2009; H10; but see S12), though the uncertainties are still too large to conclude that evolution in $\alpha$ occurs.  Robust measurements of this evolution will provide insight into hierarchical mass assembly through gas-rich mergers, and the regulation of star formation in low-mass halos.  

Of course, however, the value of $\alpha$ given by the power-law fit is an upper-limit. 
How closely this approximates the true value depends on how far the data points
are from $L^{*}$, the knee of the LF.  Given the volume probed by our observations, and previous
estimates of $L^{*}=10^{43.07\pm0.22}$ ergs s$^{-1}$ (H10; uncorrected for dust), it is clear
that the current FourStar data do not provide useful constraints on the bright, exponential end.
A complete determination of the $z=2.2$ H$\alpha$ LF from the NewH$\alpha$ survey awaits the integration
of wide-shallow data from NEWFIRM observations (C. Ly et al. 2012, in preparation).  However, 
initial comparisons can be made with the recent study of S12, 
who combine VLT/HAWK-I and UKIRT/WFCAM imaging
in the standard H$_2$ narrowband filter at 2.121 $\mu$m, and select their H$\alpha$ emitters
with a mix of photometric redshift, $BzK$, and $UBR$ selection criteria (N=45 and 518, for the
HAWK-I and WFCAM observations respectively).  The bottom panels of  
Figure~\ref{fig:lf} show that, 
despite the 
patchwork of selection methods applied by S12,
there is good agreement between
the two studies at the faint end of the fully-corrected 
LF datapoints. 
S12 report 
$\alpha=-1.57\pm0.2$ with their additional data covering higher luminosities, as they find
a relatively low $L^{*}$ of $10^{42.66\pm0.1}$ ergs s$^{-1}$ (for 0.5 mag attenuation and $\alpha$ fixed at -1.6).  If the FourStar datapoints are combined with
those from the S12 UKIRT/WFCAM imaging, the best fit Schechter parameters are  $L^{*}=10^{42.64\pm0.21}$ ergs s$^{-1}$, log($\phi^*$)=
-2.70$\pm0.30$ and $\alpha=-1.58\pm$0.40.  







\section{Summary and Future Work}

In this paper, we have presented first results from near-infrared narrowband 
imaging observations with the new FourStar camera at the 6.5m Magellan telescope.  
We have 
shown that our implementation of the dual narrowband technique for isolating 
deep intermediate redshift emission-line selected samples is highly effective, 
and identifies the $z=2.2$ H$\alpha$ emitters our NB210 excess sample 
with $\gtrsim$80\% completeness.  
The dataset is the deepest of its kind: 
the exquisite image quality ($\sim$0\farcs55 FWHM) resulting from the combination of 
instrument, telescope and site 
results in 1.19 $\mu$m and 2.10 $\mu$m imaging  
which probe 3$\sigma$ emission-line depths down to 
1.0$\times$10$^{-17}$ erg s$^{-1}$ cm$^{-2}$ 
and 1.2$\times$10$^{-17}$ erg s$^{-1}$ cm$^{-2}$
respectively, and is competitive with recent HAWK-I narrowband observations at the 8.2m VLT (e.g., Hayes et al. 2010).  Our $z=2.2$ H$\alpha$ sample extends the luminosities and SFRs probed by the well-studied UV-selected $z\sim2$ galaxies of Erb et al. (2006c) by a factor of at least two.  Optimization of the data reduction and analysis will further enable us to 
increase the depth of the sample by up to 40\%.

We compute constraints on the faint-end slope of the $z=2.2$ H$\alpha$ luminosity function
using both the dual narrowband selected H$\alpha$ emitters (N=41), and
a combined sample of dual narrowband and photometric redshift selected galaxies (N=50).
Fitting of a pure 
power-law gives an upper-limit on $\alpha$ of $-1.85\pm$0.31, which is steeper than other recent
estimates based on coarser selection techniques, but consistent
within the large uncertainties that currently characterize such
measurements.  Combining our LF points with those
at higher luminosities from other work, the slope decreases to
$\alpha=-1.58\pm0.40$.     
Additional data recently obtained from an FourStar 
pointing targeting the CANDELS area in the CDF-S will enlarge 
the sample, and tighten this constraint.

These ``narrow-deep" FourStar
observations have been obtained as part
of the larger NewH$\alpha$ Survey, which will
combine the data with ``wide-shallow" imaging through a similar
narrowband filter pair with 
NEWFIRM at the KPNO/CTIO 4m telescopes, to carry out
a statistical study of both luminous (but rare) 
and faint emission-line
galaxies in the intermediate redshift universe. The dataset, in combination with the wealth of multiwavelength observations available in the extragalactic deep fields targeted, form the basis
for follow-up studies on the physical properties (morphologies, dust, metallicities, recent star formation histories and stellar populations) of star-forming galaxies at a 
critical phase in their evolution.  The FourStar observations, in particular, enable study of lower mass members of the population, even probing SFRs down to LMC-type activities of a few tenths at $z\sim0.8$.

\acknowledgments
This project, including the purchase of the custom narrowband filters,
was made possible through the support of Hubble and Carnegie 
Fellowships awarded to JCL.  Special thanks are due to the FourStar 
instrument team and Carnegie Observatories, whose dedicated, long-term 
efforts to build an exceptional camera have enabled the science 
imagined for over a decade. 

We thank Ryan Quadri and Rik Williams for useful discussions
on using the EAZY photometric redshift code, 
and Gabe Brammer for providing the most recent version of the code.
Helpful discussions with Ranga-Ram Chary about 
the application of extinction curves, with Peter Capak about 
luminosity functions, and with Dan Kelson about
aspects of FourStar data reduction are acknowledged.



{\it Facilities:} \facility{Magellan:Baade (FourStar)}, \facility{Subaru (Suprime-Cam)} 
\clearpage




\begin{deluxetable}{cccccc}\tablecolumns{6}
\tablewidth{0pc}
\tablecaption{FourStar Observations for $\alpha=10^h00^m32^s.4$ $\delta=+02^{\circ}16^m58^s$}

\tablehead{
\colhead{Filter} &
\colhead{Int. Time} &
\colhead{FWHM} &
\colhead{3-$\sigma$ Depth}\\
&
\colhead{(hours)} &
\colhead{(\arcsec)} &
\colhead{(AB, in 1\farcs2 aperture)}\\
}
\startdata
NB119	& 9.65 & 0.56 & 25.1\\ 
NB210 	& 8.39 & 0.53 & 24.2\\
$J2$    & 7.16 & 0.61 & 26.0\\
$J3$    & 8.36 & 0.60 & 25.7\\
$K_s$ 	& 5.82 & 0.51 & 25.0\\
\enddata
\label{tab:observations}
\end{deluxetable}

\begin{deluxetable}{ccccccc}\tablecolumns{7}
\tablewidth{0pc}
\tablecaption{H$\alpha$ Luminosity Functions at $z=2.2$}

\tablehead{
\colhead{log L} &
\colhead{$\Phi(L)$} &
\colhead{N}&
\colhead{}&
\colhead{log L} &
\colhead{$\Phi(L)$} &
\colhead{N}\\
}

\startdata
\multicolumn{7}{c}{Dual emitters (N=41)}\\
\hline
\multicolumn{3}{c}{Raw}&&\multicolumn{3}{c}{Corrected}\\
\cline{1-3}
\cline{5-7}
41.77 & 6.838E-03 $\pm$ 3.419E-03 &  4 && \\
41.90 & 5.861E-03 $\pm$ 1.566E-03 & 14 && 41.94 & 8.096E-03 $\pm$ 2.649E-03 &  9 \\ 
42.10 & 4.058E-03 $\pm$ 1.223E-03 & 11 && 42.10 & 5.027E-03 $\pm$ 1.451E-03 & 12 \\
42.30 & 2.582E-03 $\pm$ 9.760E-04 &  7 && 42.30 & 4.429E-03 $\pm$ 1.278E-03 & 12 \\
42.50 & 1.476E-03 $\pm$ 7.378E-04 &  4 && 42.50 & 2.214E-03 $\pm$ 9.038E-04 &  6 \\
42.70 & 3.689E-04 $\pm$ 3.689E-04 &  1 && 42.70 & 7.378E-04 $\pm$ 5.217E-04 &  2 \\
\hline
\multicolumn{7}{c}{Dual emitters + photo-z selected (N=50)}\\
\hline
\multicolumn{3}{c}{Raw}&&\multicolumn{3}{c}{Corrected}\\
\cline{1-3}
\cline{5-7}
41.74 & 5.569E-03 $\pm$ 2.274E-03 &  6 &&  \\
41.90 & 7.966E-03 $\pm$ 1.878E-03 & 18 &&  41.93 & 1.021E-02 $\pm$ 3.005E-03 & 11 \\
42.10 & 4.427E-03 $\pm$ 1.278E-03 & 12 &&  42.10 & 6.703E-03 $\pm$ 1.675E-03 & 16 \\
42.30 & 2.951E-03 $\pm$ 1.043E-03 &  8 &&  42.30 & 4.798E-03 $\pm$ 1.330E-03 & 13 \\
42.50 & 1.844E-03 $\pm$ 8.248E-04 &  5 &&  42.50 & 2.952E-03 $\pm$ 1.044E-03 &  8 \\
42.70 & 3.689E-04 $\pm$ 3.689E-04 &  1 &&  42.70 & 7.378E-04 $\pm$ 5.217E-04 &  2 \\

\enddata
\label{tab:lf}
\end{deluxetable}

\begin{thebibliography}{}

\bibitem[Abazajian et al.(2009)]{2009ApJS..182..543A} Abazajian, K.~N., 
Adelman-McCarthy, J.~K., Ag{\"u}eros, M.~A., et al.\ 2009, \apjs, 182, 543 
\bibitem[Brammer et al.(2008)]{2008ApJ...686.1503B} Brammer, G.~B., van 
Dokkum, P.~G., \& Coppi, P.\ 2008, \apj, 686, 1503
\bibitem[Brinchmann et al.(2004)]{2004MNRAS.351.1151B} Brinchmann, J., 
Charlot, S., White, S.~D.~M., et al.\ 2004, \mnras, 351, 1151 
\bibitem[Bruzual 
\& Charlot(2003)]{2003MNRAS.344.1000B} Bruzual, G., \& Charlot, S.\ 2003, \mnras, 344, 1000  
\bibitem[Calzetti et al.(2000)]{2000ApJ...533..682C} Calzetti, D., Armus, 
L., Bohlin, R.~C., et al.\ 2000, \apj, 533, 682 
\bibitem[Daddi et al.(2004)]{2004ApJ...617..746D} Daddi, E., Cimatti, A., 
Renzini, A., et al.\ 2004, \apj, 617, 746 
\bibitem[Erb et al.(2006)]{2006ApJ...647..128E} Erb, D.~K., Steidel, C.~C., 
Shapley, A.~E., et al.\ 2006, \apj, 647, 128 
\bibitem[Erb et al.(2006)]{2006ApJ...646..107E} Erb, D.~K., Steidel, C.~C., 
Shapley, A.~E., et al.\ 2006, \apj, 646, 107 
\bibitem[Erb et al.(2006)]{2006ApJ...644..813E} Erb, D.~K., Shapley, A.~E., 
Pettini, M., et al.\ 2006, \apj, 644, 813 
\bibitem[Erben et 
al.(2009)]{2009A&A...493.1197E} Erben, T., Hildebrandt, H., Lerchster, M., et al.\ 2009, \aap, 493, 1197 
\bibitem[Fujita et al.(2003)]{2003ApJ...586L.115F} Fujita, S.~S., Ajiki, 
M., Shioya, Y., et al.\ 2003, \apjl, 586, L115 
\bibitem[Geach et al.(2008)]{2008MNRAS.388.1473G} Geach, J.~E., Smail, I., 
Best, P.~N., Kurk, J., Casali, M., Ivison, R.~J., 
\& Coppin, K.\ 2008, \mnras, 388, 1473
\bibitem[Grogin et al.(2011)]{2011ApJS..197...35G} Grogin, N.~A., Kocevski, 
D.~D., Faber, S.~M., et al.\ 2011, \apjs, 197, 35
\bibitem[Harris 
\& Zaritsky(2009)]{2009AJ....138.1243H} Harris, J., \& Zaritsky, D.\ 2009, \aj, 138, 1243   
\bibitem[Hayes et al.(2010)]{2010A&A...509L...5H} Hayes, M., Schaerer, D., Ostlin, G.\ 2010, \aap, 509, L5 
\bibitem[Hayes et al.(2010)]{2010Natur.464..562H} Hayes, M., et al.\ 2010, 
\nat, 464, 562
\bibitem[Hopkins et al.(2001)]{2001AJ....122..288H} Hopkins, A.~M., 
Connolly, A.~J., Haarsma, D.~B., \& Cram, L.~E.\ 2001, \aj, 122, 288 
\bibitem[Ilbert et al.(2009)]{2009ApJ...690.1236I} Ilbert, O., Capak, P., 
Salvato, M., et al.\ 2009, \apj, 690, 1236 
\bibitem[Jansen et al.(2001)]{2001ApJ...551..825J} Jansen, R.~A., Franx, 
M., \& Fabricant, D.\ 2001, \apj, 551, 825 
\bibitem[Kennicutt(1998)]{1998ARA&A..36..189K} Kennicutt, R.~C., Jr.\ 1998, \araa, 36, 189 
\bibitem[Kennicutt et al.(2008)]{2008ApJS..178..247K} Kennicutt, R.~C., 
Jr., Lee, J.~C., Funes, S.~J., Jos{\'e} G., Sakai, S., 
\& Akiyama, S.\ 2008, \apjs, 178, 247 
\bibitem[Kodama et al.(2004)]{2004MNRAS.354.1103K} Kodama, T., Balogh, 
M.~L., Smail, I., Bower, R.~G., \& Nakata, F.\ 2004, \mnras, 354, 1103 
\bibitem[Koekemoer et al.(2011)]{2011ApJS..197...36K} Koekemoer, A.~M., 
Faber, S.~M., Ferguson, H.~C., et al.\ 2011, \apjs, 197, 36  
\bibitem[Labb{\'e} et al.(2003)]{2003AJ....125.1107L} Labb{\'e}, I., Franx, 
M., Rudnick, G., et al.\ 2003, \aj, 125, 1107
\bibitem[Lee et al.(2009)]{2009ApJ...706..599L} Lee, J.~C., Gil de Paz, A., 
Tremonti, C., et al.\ 2009, \apj, 706, 599 
\bibitem[Lee et al.(2007)]{2007ApJ...671L.113L} Lee, J.~C., Kennicutt, 
R.~C., Funes, S.~J., Jos{\'e} G., Sakai, S., \& Akiyama, S.\ 2007, \apjl, 671, L113 
\bibitem[Lilly et al.(2007)]{2007ApJS..172...70L} Lilly, S.~J., Le 
F{\`e}vre, O., Renzini, A., et al.\ 2007, \apjs, 172, 70  
\bibitem[Ly et al.(2011)]{2011ApJ...726..109L} Ly, C., Lee, J.~C., Dale, 
D.~A., Momcheva, I., Salim, S., Staudaher, S., Moore, C.~A., 
\& Finn, R.\ 2011, \apj, 726, 109
\bibitem[Mannucci et al.(2010)]{2010MNRAS.408.2115M} Mannucci, F., Cresci, 
G., Maiolino, R., Marconi, A., \& Gnerucci, A.\ 2010, \mnras, 408, 2115 
\bibitem[McCracken et al.(2010)]{2010ApJ...708..202M} McCracken, H.~J., 
Capak, P., Salvato, M., et al.\ 2010, \apj, 708, 202  
\bibitem[Moustakas et al.(2006)]{2006ApJ...642..775M} Moustakas, J., 
Kennicutt, R.~C., Jr., \& Tremonti, C.~A.\ 2006, \apj, 642, 775 
\bibitem[Nakajima et al.(2012)]{2012ApJ...745...12N} Nakajima, K., Ouchi, 
M., Shimasaku, K., et al.\ 2012, \apj, 745, 12
\bibitem[Noeske et al.(2007)]{2007ApJ...660L..47N} Noeske, K.~G., Faber, 
S.~M., Weiner, B.~J., et al.\ 2007, \apjl, 660, L47 
\bibitem[Ono et al.(2012)]{2012ApJ...744...83O} Ono, Y., Ouchi, M., 
Mobasher, B., et al.\ 2012, \apj, 744, 83 
\bibitem[Ouchi et al.(2010)]{2010ApJ...723..869O} Ouchi, M., Shimasaku, K., 
Furusawa, H., et al.\ 2010, \apj, 723, 869  
\bibitem[Nilsson et 
al.(2007)]{2007A&A...474..385N} Nilsson, K.~K., Orsi, A., Lacey, C.~G., Baugh, C.~M., \& Thommes, E.\ 2007, \aap, 474, 385 
\bibitem[Reddy et al.(2010)]{2010ApJ...712.1070R} Reddy, N.~A., Erb, D.~K., 
Pettini, M., Steidel, C.~C., \& Shapley, A.~E.\ 2010, \apj, 712, 1070 
\bibitem[Reddy et al.(2008)]{2008ApJS..175...48R} Reddy, N.~A., Steidel, 
C.~C., Pettini, M., et al.\ 2008, \apjs, 175, 48 
\bibitem[Schechter(1976)]{1976ApJ...203..297S} Schechter, P.\ 1976, \apj, 
203, 297
\bibitem[Shioya et al.(2008)]{2008ApJS..175..128S} Shioya, Y., Taniguchi, 
Y., Sasaki, S.~S., et al.\ 2008, \apjs, 175, 128 
\bibitem[Spitler et al.(2012)]{2012ApJ...748L..21S} Spitler, L.~R., 
Labb{\'e}, I., Glazebrook, K., et al.\ 2012, \apjl, 748, L21 
\bibitem[Sobral et al.(2012)]{2012MNRAS.420.1926S} Sobral, D., Best, P.~N., 
Matsuda, Y., et al.\ 2012, \mnras, 420, 1926 
\bibitem[Sobral et al.(2009)]{2009MNRAS.398...75S} Sobral, D., Best, P.~N., 
Geach, J.~E., et al.\ 2009, \mnras, 398, 75
\bibitem[Scoville et al.(2007)]{2007ApJS..172....1S} Scoville, N., Aussel, 
H., Brusa, M., et al.\ 2007, \apjs, 172, 1 
\bibitem[Taniguchi et al.(2007)]{2007ApJS..172....9T} Taniguchi, Y., 
Scoville, N., Murayama, T., et al.\ 2007, \apjs, 172, 9 
\bibitem[Taniguchi et al.(2003)]{2003JKAS...36..123T} Taniguchi, Y., 
Shioya, Y., Ajiki, M., et al.\ 2003, Journal of Korean Astronomical 
Society, 36, 123 
\bibitem[Teplitz et al.(1998)]{1998ApJ...506..519T} Teplitz, H.~I., Malkan, 
M., \& McLean, I.~S.\ 1998, \apj, 506, 519 
\bibitem[Tresse et al.(2002)]{2002MNRAS.337..369T} Tresse, L., Maddox, 
S.~J., Le F{\`e}vre, O., \& Cuby, J.-G.\ 2002, \mnras, 337, 369  
\bibitem[van Dokkum et al.(2009)]{2009PASP..121....2V} van Dokkum, P.~G., 
et al.\ 2009, \pasp, 121, 2
\bibitem[Villar et al.(2008)]{2008ApJ...677..169V} Villar, V., Gallego, J., 
P{\'e}rez-Gonz{\'a}lez, P.~G., et al.\ 2008, \apj, 677, 169  
\bibitem[Whitaker et al.(2011)]{2011ApJ...735...86W} Whitaker, K.~E., et 
al.\ 2011, \apj, 735, 86
\bibitem[Whitney et al.(2008)]{2008AJ....136...18W} Whitney, B.~A., Sewilo, 
M., Indebetouw, R., et al.\ 2008, \aj, 136, 18 

\end{thebibliography}
\end{document}